\documentclass[%
 reprint,
 amsmath,amssymb,
 aps,
]{revtex4-2}

\usepackage{graphicx}% Include figure files
\usepackage{dcolumn}% Align table columns on decimal point
\usepackage{upgreek}
\usepackage{threeparttable}
\usepackage{bm} % for \boldsymbol
\usepackage[colorlinks, allcolors=blue]{hyperref}% add hypertext capabilities

\begin{document}
\preprint{APS/123-QED}

\title{Thickness bound for nonlocal wide-field-of-view metalenses}

\author{Shiyu Li}

\affiliation{Ming Hsieh Department of Electrical and Computer Engineering, University of Southern California, Los Angeles, California 90089, USA}

\author{Chia Wei Hsu}
\email{cwhsu@usc.edu}
\affiliation{Ming Hsieh Department of Electrical and Computer Engineering, University of Southern California, Los Angeles, California 90089, USA}

%% To be edited by editor
% \dates{Compiled \today}

%\ociscodes{(140.3490) Lasers, distributed feedback; (060.2420) Fibers, polarization-maintaining;(060.3735) Fiber Bragg gratings.}

%% To be edited by editor
% \doi{\url{http://dx.doi.org/10.1364/XX.XX.XXXXXX}}

% LSA abstract length limit: 250 words
% Nature Communications length limit: 150 words
% Optica abstract length: approximately 100 words (not strict)
% Current number of words: 182
\begin{abstract}
Metalenses---flat lenses made with optical metasurfaces---promise to enable thinner, cheaper, and better imaging systems.
Achieving a sufficient angular field of view (FOV) is crucial toward that goal and requires a tailored incident-angle-dependent response. %the subject of much recent work.
Here, we show that there is an intrinsic trade-off between achieving a desired broad-angle response and reducing the thickness of the device. It originates from the Fourier transform duality between space and angle.
%, which is applicable to any structural design and any material composition. Based on the properties of ideal wide-FOV lenses, we build their transmission matrices, quantify their degree of nonlocality, from which we determine their minimal thickness. 
One can write down the transmission matrix describing the desired angle-dependent response, convert it to the spatial basis where its degree of nonlocality can be quantified through a lateral spreading, and determine the minimal device thickness based on such a required lateral spreading.
This approach is general.
When applied to wide-FOV lenses, it predicts the minimal thickness as a function of the FOV, lens diameter, and numerical aperture.
The bound is tight, as some inverse-designed multi-layer metasurfaces can approach the minimal thickness we found.
%, while existing metalenses based on doublets or aperture stops are one-to-two-orders-of-magnitude thicker.
%Our transmission-matrix approach can also determine the thickness bounds of other systems beyond lenses.
This work offers guidance for the design of nonlocal metasurfaces, proposes a new framework for establishing bounds, and reveals the relation between angular diversity and spatial footprint in multi-channel systems.
\end{abstract}

%\setboolean{displaycopyright}{true}

%\begin{document}

\maketitle

\begin{figure*}[t]
\centering
\includegraphics[width=0.85\textwidth]{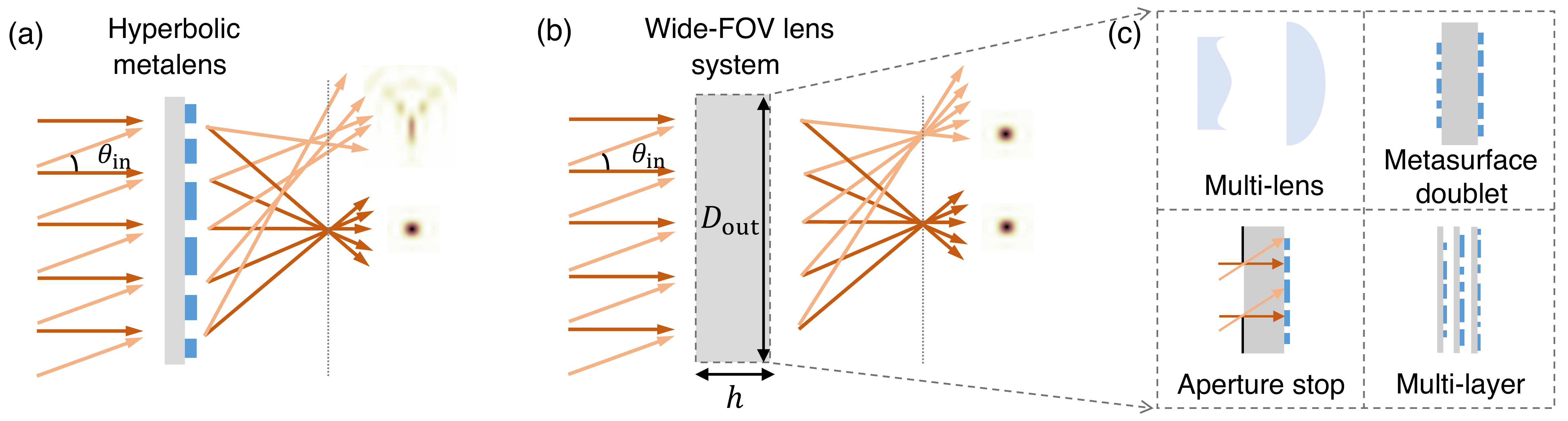}
\caption{Wide-FOV lens systems. (a,b) Schematics of (a) a metalens with a hyperbolic phase profile and (b) a diffraction-limited lens system with a wide FOV. The former can have subwavelength thickness and produces a diffraction-limited focal spot at normal incidence but suffers from strong aberrations at oblique incidence. The latter achieves diffraction-limited focusing over a wide range of incident angles but requires a minimal thickness $h$. 
%The black dotted line denotes the imaging surface. The diameter $D$ and thickness $h$ of the lens system are marked.
(c) Examples of systems that realize wide-FOV diffraction-limited focusing: cascade of multiple lenses, metasurface doublets, use of an aperture stop, and multi-layer metasurfaces.}
\label{fig:fig1}
\end{figure*}

%\section{Introduction}
Metasurfaces use subwavelength building blocks to achieve versatile functions with spatially-resolved modulation of the phase, amplitude, and polarization of light~\cite{yu2011light,kildishev2013planar,yu2014flat,chen2016review,genevet2017recent,hsiao2017fundamentals,kamali2018review,chen2020flat}. Among them, metalenses~\cite{lalanne2017metalenses,khorasaninejad2017metalenses,tseng2018metalenses,liang2019high,kim2021dielectric} receive great attention given their potential to enable thinner, lighter, cheaper, and better imaging systems for a wide range of applications where miniaturization is critical ({\it e.g.} for bio-imaging and endoscopy and for mobile and wearable devices such as cell phones and mixed-reality headsets).
Metalenses are commonly modeled by a spatially-varying transmission phase-shift profile $\phi(x,y)$ where $x,y$ are the transverse coordinates. To focus normal-incident light to a diffraction-limited spot with focal length $f$, one can require all of the transmitted light to be in phase when reaching the focal spot, which gives a hyperbolic phase profile~\cite{hecht2017optics,aieta2012aberration}
\begin{equation}
\phi_{\rm hyp}(x,y)=\frac{2\pi}{\lambda} \left (f-\sqrt{f^2+x^2+y^2} \right )
\label{eq:hyp_phase}
\end{equation}
where $\lambda$ is the operating wavelength. However, for oblique illumination, the optical path lengths of the marginal rays no longer match that of the chief ray, resulting in coma, astigmatism, and field-curvature aberrations~\cite{aieta2013aberrations,kalvach2016aberration,decker2019imaging} as schematically illustrated in Fig.~\ref{fig:fig1}(a). These aberrations severely limit the input angular range over which focusing is achieved ({\it i.e.}, the FOV). %, even though FOV is important for most applications.

One way to expand the FOV is to use the phase profile of an equivalent spherical lens~\cite{liang2019high} or a quadratic phase profile~\cite{pu2017nanoapertures,martins2020metalenses,lassalle2021imaging}, which reduce off-axis aberrations. However, doing so introduces spherical aberration and defocus aberration, with a reduced effective aperture size, axial elongation, and a low Strehl ratio~\cite{liang2019high,lassalle2021imaging,2022_Lin_arXiv}, so the focus is no longer diffraction-limited.
%When pushing the radius and refractive index of spherical lenses to infinity, a widely used angle-independent phase distribution, the so-called parabolic (quadratic) phase, can be obtained to realize a large FOV \cite{pu2017nanoapertures,martins2020metalenses}.

To achieve wide FOV with diffraction-limited focusing, one can use metasurface doublets~\cite{arbabi2016miniature,groever2017meta,he2019polarization,li2021super,tang2020achromatic,kim2020doublet,huang2021achromatic,martins2022fundamental} or triplets~\cite{shrestha2019multi} analogous to conventional multi-lens systems, add an aperture stop so incident light from different angles reach different regions of the metasurface~\cite{engelberg2020near,shalaginov2020single,fan2020ultrawide,zhang2021extreme,yang2021design,yang2021wide}, or use inverse-designed multi-layer structures~\cite{lin2018topology,lin2021computational};
these approaches are schematically illustrated in Fig.~\ref{fig:fig1}(b,c).
%Another method suggests patterning metasurfaces on spherical surfaces \cite{aieta2013aberrations}, which poses a challenge on fabrication.
Notably, all of these approaches involve a much thicker system where the overall thickness ({\it e.g.}, separation between the aperture stop and the metasurface) plays a critical role.
Meanwhile, miniaturization is an important consideration and motivation for metalenses.
This points to the scientifically and technologically important questions: is there a fundamental trade-off between the FOV and the thickness of a metalens system, or lenses in general? If so, what is the minimal thickness allowed by physical laws?

Light propagating through disordered media exhibits an angular correlation called ``memory effect''~\cite{freund1988memory,berkovits1989memory,osnabrugge2017generalized,yilmaz2019angular,yilmaz2021customizing}: when the incident angle tilts, the transmitted wavefront stays invariant and tilts by the same amount if the input momentum tilt is smaller than roughly one over the medium thickness. 
Weakly scattering media like a diffuser exhibit a longer memory effect range~\cite{2015_Schott_OE}, and thin layers like a metasurface also have a long memory effects range~\cite{jang2018wavefront}.
With angle-multiplexed volume holograms, it was found that a thicker hologram material is needed to store more pages of information at different angles~\cite{li1994three,barbastathis2000volume}.
These phenomena suggest there may be an intrinsic relation between angular diversity and thickness in multi-channel systems including but not limited to lenses.

Bounds for metasurfaces can provide valuable physical insights and guidance for future designs.
%Aieta \emph{et al.} \cite{aieta2013aberrations} showed that the correction of the spherical aberration for on-axis incidence leads to the degradation of focusing for off-axis inputs with a single metasurface.
%Liang \emph{et al.} \cite{liang2019high} observed a trade-off between NA and FOV, highlighting the difficulty of correcting both spherical and off-axis aberrations simultaneously for metalenses with large NA.
Shrestha \emph{et al.}~\cite{shrestha2018broadband} and Presutti \emph{et al.}~\cite{presutti2020focusing} related the maximal operational bandwidth of achromatic metalenses to the numerical aperture (NA), lens diameter, and thickness, %They also briefly demonstrated the trade-off between the achievable bandwidth and the transmission efficiency due to their opposite responses with respect to the thickness changes.
which was generalized to wide-FOV operation by Shastri \emph{et al.}~\cite{shastri2022bandwidth} and diffractive lenses by Engelberg \emph{et al.}~\cite{engelberg2021achromatic}.
%showed the relation between the spectral range and Fresnel number for achromatic metalenses composed of truncated waveguide nanopillars %found the relation between the spectral range, Fresnel number, nanostructure height and its refractive index contrast with respect to its surrounding for achromatic metalenses composed of truncated waveguide nanopillars.
%Zhang \emph{et al}.\ established bounds on power concentration~\cite{zhang2019scattering}, 
Shastri \emph{et al}.\ investigated the relation between absorber efficiency and its omnidirectionality~\cite{shestri2021existence}, Gigli \emph{et al}.\ analyzed the limitations of Huygens' metasurfaces due to nonlocal interactions~\cite{gigli2021fundamental}, Chung \emph{et al}.\ determined the upper bounds on the efficiencies of unit-cell-based high-NA metalenses~\cite{chung2020high},
%Other trade-offs for different types of metasurfaces have also been discussed, such as metasurfaces governed by ray optics \cite{estakhri2016wave}, with ultrathin footprint \cite{arbabi2017fundamental}.
Yang \emph{et al}.\ quantified the relation between optical performance and design parameters for aperture-stop-based metalenses~\cite{yang2021wide}, and Martins \emph{et al}.\ studied the trade-off between the resolution and FOV for doublet-based metalenses~\cite{martins2022fundamental}.
Each of these studies concerns one specific type of design.
The power-concentration bound of Zhang \emph{et al}~\cite{zhang2019scattering}
and the multi-functional bound of Shim \emph{et al}~\cite{shim2021fundamental} are more general, though they bound the performance rather than the device footprint.
However, the relationship between thickness and angular diversity remains unknown.

In this work, we establish such relationship and apply it to wide-FOV metalenses. Given any desired angle-dependent response, we can write down its transmission matrix, measure its degree of nonlocality (as encapsulated in the lateral spreading of incident waves encoded in the transmission matrix), from which we determine the minimal device thickness. %We also map out how such minimal lens thickness depends on the FOV, diameter of the output aperture, and NA. 
This is a new approach for establishing bounds, applicable across different designs including single-layer metasurfaces, cascaded metasurfaces, diffractive lenses, bulk metamaterials, thick volumetric structures, {\it etc}. %as well as designs not based on metasurfaces.
%Some inverse-designed multi-layer metasurfaces~\cite{lin2018topology,lin2021computational} can approach this bound, indicating the bound is tight.
%Our work provides guidance on the design of wide-FOV metalens systems.
%The transmission-matrix approach can also be used to establish the thickness bounds and other properties of nonlocal metasurfaces beyond wide-FOV lenses.

\vspace{-1pt}
%\section{Results}
\section{Thickness bound via transmission matrix}

%We look for universal bounds applicable to all designs, including those illustrated in Fig.~\ref{fig:fig1}(c) and still-unknown ones. To do so, we adopt the transmission matrix formalism.
The multi-channel transport through any linear system can be described by a transmission matrix.
Consider monochromatic wave at angular frequency $\omega = 2\pi c/\lambda$.
The incoming wavefront can be written as a superposition of propagating waves at different angles and polarizations, as
\begin{equation}
%{\bf E}_{\rm in}({\boldsymbol \rho},z=0)=\sum_{a} {\bf E}_{\rm in}({\bf k}_{\parallel}^a,z=0)w_R(\boldsymbol \rho)e^{i{\bf k}_{\parallel}^a\cdot {\boldsymbol \rho}}\Delta{\bf k}_{\parallel}^a,
{\bf E}_{\rm in}({\boldsymbol \rho},z=0)=\sum_{a=1}^{N_{\rm in}} v_a \hat{e}_a e^{i{\bf k}_{\parallel}^a\cdot {\boldsymbol \rho}} w_{\rm in}(\boldsymbol \rho), 
\label{eq:E_in}
\end{equation}
where ${\boldsymbol \rho}=(x,y)$ is the transverse coordinate; $\hat{e}_a$ and ${\bf k}_{\parallel}^a = (k_x^a, k_y^a)$ are the polarization state and the transverse wave number (momentum) of the $a$-th plane-wave input with amplitude $v_a$; $z=0$ is the front surface of the lens, and $w_{\rm in}(\boldsymbol \rho)=1$ for $|\boldsymbol \rho|<D_{\rm in}/2$ (zero otherwise) is a window function that describes an aperture that blocks incident light beyond entrance diameter $D_{\rm in}$.
The wave number ${\bf k}_{\parallel}^a$ is restricted to propagating waves within the angular FOV, with $|{\bf k}_{\parallel}^a|<(\omega/c) \sin({\rm FOV/2})$. % where $n_{\rm in}$ is the refractive index of the substrate on the incident side.
%There is typically an entrance aperture that blocks incident light beyond diameter $D_{\rm in}$, so ${\bf E}_{\rm in}({\boldsymbol \rho},z=0) = 0$ for $|{\boldsymbol \rho}| > D_{\rm in}/2$.
Since the input is band-limited in space due to the entrance aperture, a discrete sampling of ${\bf k}_{\parallel}^a$ with $2\pi/D_{\rm in}$ spacing at the Nyquist rate~\cite{landau1967sampling} is sufficient. %, allowing Eq.~\eqref{eq:E_in} to be written as a discrete summation over a finite number of incident wavevectors and the two polarization components, with
Therefore, the number $N_{\rm in}$ of ``input channels'' is finite~\cite{miller2019waves},
and the incident wavefront is parameterized by a column vector ${\boldsymbol v} = [v_1,\cdots,v_{N_{\rm in}}]^{\rm T}$.
Similarly, the propagating part of the transmitted wave is a superposition of $N_{\rm out}$ output channels at different angles and polarizations,
\begin{equation}
{\bf E}_{\rm t}({\boldsymbol \rho},z=h)=\sum_{b=1}^{N_{\rm out}} u_b \hat{e}_b e^{i{\bf k}_{\parallel}^b\cdot {\boldsymbol \rho}} w_{\rm out}(\boldsymbol \rho),
\label{eq:E_out}
\end{equation}
where $h$ is the thickness of the lens system, and the window function $w_{\rm out}(\boldsymbol \rho)=1$ for $|\boldsymbol \rho|<D_{\rm out}/2$ blocks transmitted light beyond an output aperture with diameter $D_{\rm out}$.
The transmitted wavefront is parameterized by column vector ${\boldsymbol u} = [u_1,\cdots,u_{N_{\rm out}}]^{\rm T}$.
Normalization prefactors are ignored in Eqs.~\eqref{eq:E_in}--\eqref{eq:E_out} for simplicity.

\begin{figure*}[tb]
\centering
\includegraphics[width=1.0\textwidth]{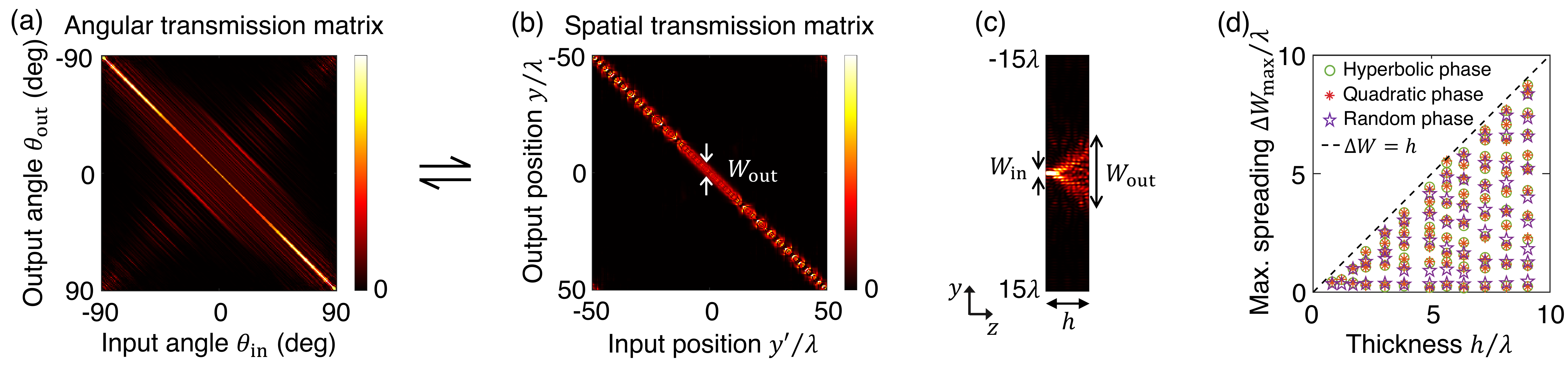}
\caption{Transmission matrix and its relation to nonlocality and device thickness $h$. (a,b) The same transmission matrix in angular basis $|t(k_y,k_y')|^2$ and in spatial basis $|t(y,y')|^2$, for a hyperbolic metalens with diameter $D=100\lambda$, NA = 0.45, thickness $h=4.2\lambda$, index contrast = 0.4, without a substrate ($n_{\rm in}=1$).
The axes in (a) are linearly spaced in $k_y=(\omega/c)\sin{\theta_{\rm out}}$ and $k_y'=(\omega/c)\sin{\theta_{\rm in}}$.
(c) Intensity profile inside the metasurface for a localized input at $y'=0$, corresponding to the middle column of the spatial transmission matrix. The lateral spreading $\Delta W=W_{\rm out}-W_{\rm in}$ quantifies the degree of nonlocality. (d) Maximal lateral spreading $\Delta W_{\rm max} \equiv \max_{y'} \Delta W(y')$ computed from $t(y,y')$, for random metasurfaces and standard metasurfaces with varying parameters and thicknesses. The data reveal an empirical inequality $\Delta W_{\rm max} < h$, which places a lower bound on the thickness of a device that realizes the corresponding transmission matrix.}
\label{fig:fig2}
\end{figure*}

The input and the output must be related through a linear transformation, so we can write
\begin{equation}
u_b = \sum_{a=1}^{N_{\rm in}} t_{ba} v_a,
\end{equation}
or ${\boldsymbol u} = {\bf t} {\boldsymbol v}$, where ${\bf t}$ is the
%Such a linear transformation is the “\emph{transmission matrix}” ${\bf t}({\bf k}_{\parallel},{\bf k}_{\parallel}^{'})$, such that ${\boldsymbol u}= {\bf t}({\bf k}_{\parallel},{\bf k}_{\parallel}^{'}){\boldsymbol v}$, namely $u_b = \sum_a t_{ba} v_a$ with $t_{ba}={\bf t}({\bf k}_{\parallel}^b,{\bf k}_{\parallel}^a)$. Conceptually, we can interpret the transmission matrix as
%\begin{equation}
%{\bf E}_{\rm t}({\bf k}_{\parallel}^b,z=h)=\sum_a {\bf t}({\bf k}_{\parallel}^b,{\bf k}_{\parallel}^a){\bf E}_{\rm in}({\bf k}_{\parallel}^a,z=0), %\Delta{\bf k}_{\parallel}^a,
%\label{eq:3}
%\end{equation}
%where ${\bf k}_{\parallel}^a$ and ${\bf k}_{\parallel}^b$ represent the input and output channels, respectively.
transmission matrix~\cite{popoff2010measuring,mosk2012controlling,rotter2017light}.
The transmission matrix describes the exact wave transport through any linear system, regardless of the complexity of the structure and its material compositions.

For simplicity, in the examples below we consider the transverse magnetic (TM) waves of 2D systems where we only need to consider the $\hat{x}$ polarization ${\bf E}=E_x(y,z)\hat{x}$, with the transverse coordinate $\rho=y$ and the transverse momentum $k_y$ both being scalars.
%Note that we use $\theta_{\rm in}$ (and subsequently the FOV) to denote the incident angle from air, not angle in the substrate.
%The truncated plane-wave basis is also built to be flux orthogonal, with derivations shown in Supplementary Sec.~2.
We compute the transmission matrix with full-wave simulations using the recently proposed augmented partial factorization method~\cite{2022_Lin_arXiv} implemented in the open-source software MESTI~\cite{MESTI}.
Figure~\ref{fig:fig2}(a) shows the squared amplitude of the transmission matrix for a 2D metalens designed to exhibit the hyperbolic phase profile in Eq.~\eqref{eq:hyp_phase} at normal incidence.
%Here, $\theta_{\rm in}=\arcsin{[k_y'/(n_{\rm in}\omega/c)]}$ and $\theta_{\rm out}=\arcsin{[k_y/(n_{\rm out}\omega/c)]}$ are the input and output angles.
We informally express such transmission matrix in angular basis as $t(k_y,k_y')$ where $k_y'=k_y^a=(\omega/c)\sin\theta_{\rm in}$ is the transverse momentum of the input and $k_y=k_y^b=(\omega/c)\sin\theta_{\rm out}$ is that of the output.

Each windowed plane-wave input or output is itself a superposition of spatially-localized waves, so we can convert the transmission matrix from the angular basis to a spatial basis with no change in its information content.
Informally, such a change of basis is described by a Fourier transform $F$ on the input side and an inverse Fourier transform $F^{-1}$ on the output side~\cite{judkewitz2015translation}, as
%We can use discrete Fourier transform $F$ and $F'$ to convert the standard transmission matrix in momentum basis to a transmission matrix in spatial basis
%~\cite{judkewitz2015translation,yilmaz2019transverse},
\begin{equation}
t(y,y')=F^{-1}t(k_y,k_y')F.
\label{eq:t_change_basis}
\end{equation}
%with the information content unchanged. Generally $F$ and $F'$ are different due to the different input and output channels. Note that additional prefactors are needed in Eq.~\eqref{eq:t_change_basis} to properly normalize the input and output flux, and the full expression is shown in Eq.~(\ref{eq:t_spa}) and Supplementary Sec.~2.
A formal derivation is provided in the Supplementary Materials.
Intuitively, $t(y,y')$ gives the output at position $y$ given a localized incident wave focused at $y'$; it has also been called the ``discrete-space impulse response''~\cite{torfeh2020modeling}. Its off-diagonal elements capture nonlocal couplings between different elements of a metasurface, which are commonly ignored in conventional metasurface designs but play a critical role for angular diversity.
%Note that such basis change doesn't require having access to the full $N_{\rm out}\times N_{\rm in}$ transmission matrix; we can, for example, restrict ${\bf t}({\bf k}_{\parallel},{\bf k}_{\parallel}')$ in Eq.~\eqref{eq:t_change_basis} to input channels ${\bf k}_{\parallel}'$ within the FOV of interest while considering all propagating output ${\bf k}_{\parallel}$.
%Each column of ${\bf t}({\boldsymbol \rho},{\boldsymbol \rho}')$ then provides the output when the incident wavefront consists of $|{\bf k}_{\parallel}'| < (2\pi/\lambda)\sin({\rm FOV/2})$, making up a sinc profile in 2D or a jinc ({\it i.e.}, airy-disk) profile in 3D.
%with characteristic width $\frac{\lambda/n_{\rm in}}{2\sin({\rm FOV/2})}$.
Figure~\ref{fig:fig2}(b) shows the transmission matrix of Fig.~\ref{fig:fig2}(a) in spatial basis.

The spatial transmission matrix $t(y,y')$ provides a measure of the nonlocality and an intuitive link to the system thickness $h$. It is expected that given a thicker device, incident light at $z=0$ can potentially spread more laterally when it reaches the other side at $z=h$. The extent of such a lateral spreading $\Delta W$ is the difference between the width of the output and that of the input,
\begin{equation}
\Delta W(y')=W_{\rm out}(y')-W_{\rm in},
\label{eq:lateral_spreading}
\end{equation}
as indicated in Fig.~\ref{fig:fig2}(c) on a numerically computed intensity profile with a localized incident wave. % given a sinc incident wavefront. % adopted in the spatial transmission matrix.
Such a lateral spreading $\Delta W$ quantifies the strength of nonlocal couplings.
The output width $W_{\rm out}$ is also the vertical width of the near-diagonal elements of the spatial transmission matrix $t(y,y')$, %~\cite{note_on_interpolation}, 
as indicated in Fig.~\ref{fig:fig2}(b).

To quantify the transverse widths, we use the inverse participation ratio (IPR)~\cite{yilmaz2019transverse}, with
\begin{equation}
W_{\rm out}(y')=\frac{\left [\int \left |t(y,y') \right |^2dy \right ]^2}{\int \left |t(y,y') \right |^4dy}.
\label{eq:W_out}
\end{equation}
For rectangular functions, the IPR equals the width of the function.
The width of the input is similarly defined:
in the spatial basis, each input consists of plane waves with momenta $|k_y'| < (\omega/c)\sin({\rm FOV/2})$ that make up a sinc profile in space, whose IPR is $W_{\rm in}=3\lambda/[4\sin({\rm FOV/2})]$.

%For flat structures like metasurfaces, the characteristic thickness is uniform across the structure, and $\Delta W(y')$ is weakly dependent on $y'$.
The nonlocal lateral spreading $\Delta W(y')$ depends on the location $y'$ of illumination.
Since we want to relate lateral spreading to the device footprint which is typically measured by the thickness at its thickest part, below we will consider the maximal lateral spreading across the surface,
\begin{equation}
\Delta W_{\rm max} \equiv \max_{y'} \Delta W(y').
\label{eq:dW_max}
\end{equation}

Figure~\ref{fig:fig2}(d) shows the maximal spreading $\Delta W_{\rm max}$ as a function of thickness $h$, calculated from full-wave simulations using MESTI~\cite{MESTI}.
Here we consider metasurfaces designed to have the hyperbolic phase profile of Eq.~\eqref{eq:hyp_phase} at normal incidence, metasurfaces with a quadratic phase profile~\cite{pu2017nanoapertures,martins2020metalenses,lassalle2021imaging} at normal incidence, and metasurfaces with random phase profiles. These data points cover NA from 0.1 to 0.9, index contrasts from 0.1 to 2, using diameter $D_{\rm out}=100\lambda$, with the full FOV $=180^{\circ}$.
%As mentioned above, each column of the spatial transmission matrix has a sinc illumination profile with characteristic width $\frac{3\lambda}{4n_{\rm in}}$, so the diagonal width maps to the lateral spreading only when ${\rm DW} \gg \lambda/2 n_{\rm in}$; therefore, we should ignore data points in the shaded region where ${\rm DW} \lesssim 3 \lambda/2 n_{\rm in}$.
From these data, we observe an empirical inequality
\begin{equation}
%h \gtrsim \Delta W,
\Delta W_{\rm max} < h,
\label{eq:h_and_dW}
\end{equation}
as intuitively expected. This relation provides a quantitative link between the angle-dependent response of a system and its thickness.

Note that while higher index contrasts allow a $2\pi$ phase shift to be realized with thinner metasurfaces, such higher index contrasts do not lower the minimum thickness governed by Eq.~\eqref{eq:h_and_dW}.
The systems considered in Fig.~\ref{fig:fig2}(d) have no substrate and use the full FOV; Figures~S1--S2 in the Supplementary Materials further show that Eq.~\eqref{eq:h_and_dW} also holds for metasurfaces sitting on a substrate or when a reduced FOV is considered.
In 2D, the ratio between minimal thickness and our definition of the lateral spreading happens to be roughly 1, independent of index contrast and substrate index; we expect a similar relation in 3D, likely with a slightly different ratio.

%Since the off-diagonal elements of the spatial transmission matrix $t(y,y')$ capture nonlocal effects, the lateral spreading $\Delta W$ is also a measure of the nonlocality.
%Therefore, Eq.~\eqref{eq:h_and_dW} indicates that broad-angle nonlocal responses require the system to have a minimal thickness.

We emphasize that even though Eq.~\eqref{eq:h_and_dW} follows intuition and is found to be valid across a wide range of systems considered above, it remains empirical.
In particular, in the presence of guided resonances~\cite{2002_Fan_PRB,2016_Gao_srep}, it is possible for the incident wave from free space to be partially converted to a guided wave and then radiate out to the free space after some in-plane propagation, enabling the lateral spreading $\Delta W$ to exceed the thickness $h$;
this is likely the case with resonance-based space-squeezing systems~\cite{reshef2021optic,guo2020squeeze,chen2021}.
Indeed, we have found that Eq.~\eqref{eq:h_and_dW} may be violated within a narrow angular range near that of a guided resonance.
It is possible to extend the angular range by stacking multiple resonances~\cite{chen2021} or by using guided resonances on a flat band~\cite{2018_Leykam_APX,2018_Leykam_APLPh}, but doing so restricts the degrees of freedom for further designs. % and are unlikely to be compatible with the requirement of a device like a wide-FOV metalens.
In the following, we assume Eq.~\eqref{eq:h_and_dW} is valid, which implicitly excludes broad-angle resonant effects.

%However, most guided resonances can only be excited within a narrow angular range.
%We found that Eq.~\eqref{eq:h_and_dW} holds even in the presence of guided resonances as long as the FOV is sufficiently large ({\it e.g.}, FOV $>10^{\circ}$).

Given the angle-dependent response of a system described by $t(k_y,k_y')$, Eqs.~\eqref{eq:t_change_basis}--\eqref{eq:h_and_dW} quantify its degree of nonlocality and the minimal thickness such a system must have.
This formalism applies to different nonlocal systems.
Below, we use this formalism to establish a thickness bound for wide-FOV lenses.

%Some parameters in the transmission matrix do not affect the transport properties of interest.
%To establish the smallest possible thickness $h$, we next optimize such parameters to minimize $\Delta W_{\rm max}$. 

% we can optionally comment on the squeeze-spaced papers here

\section{Thickness bound for wide-FOV lenses}
\subsection{Transmission matrix of an ideal wide-FOV lens}
\label{sec:DW}

\begin{figure}[tb]
\centering
\includegraphics[width=0.48\textwidth]{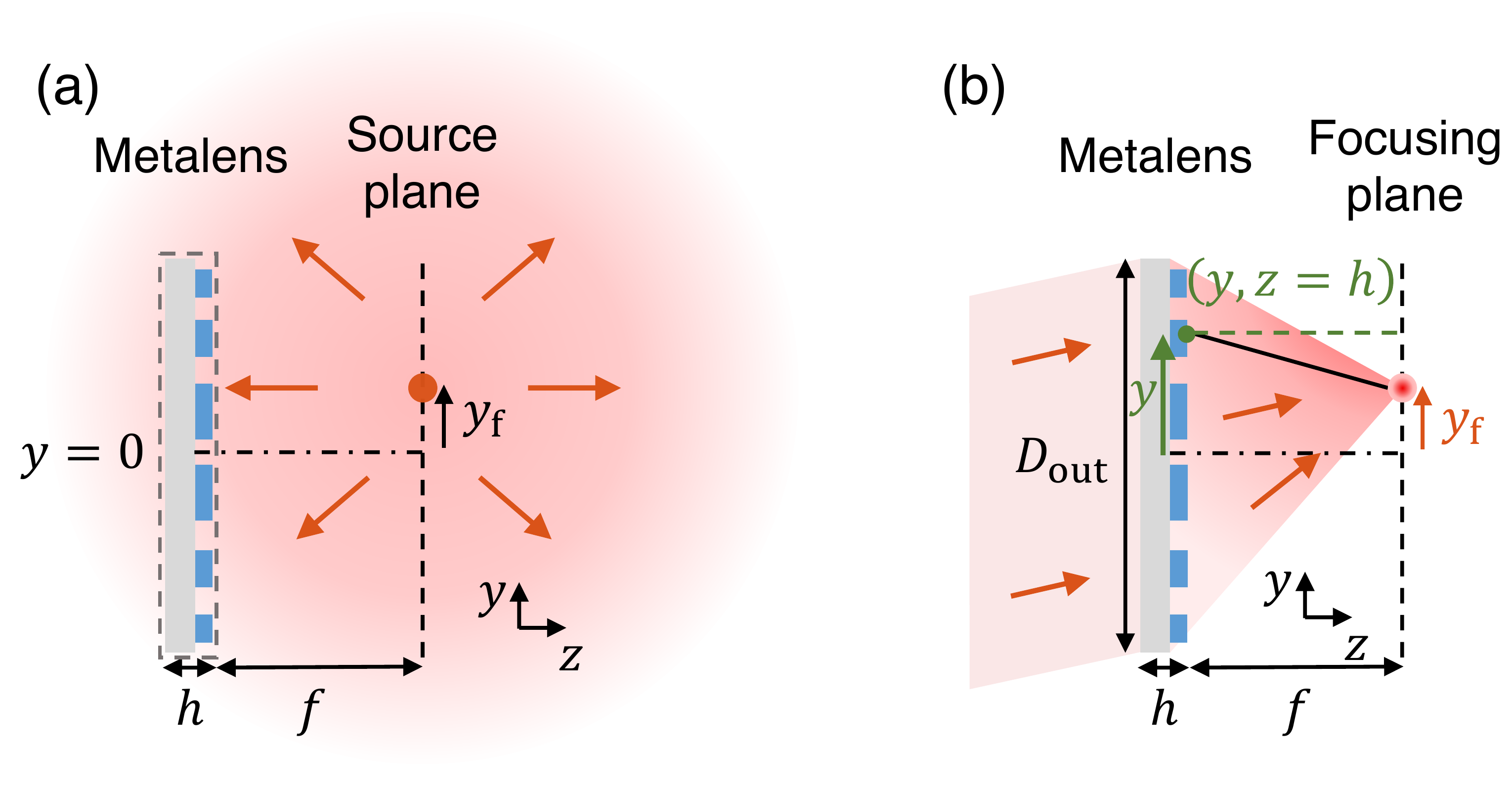}
\caption{Schematics for determining the field transmitted through an ideal metalens. (a) Outgoing field from a point source located at the focal spot $(y=y_{\rm f},z=h+f)$. (b) To ideally focus light to such focal spot, the transmitted field across the back aperture $D_{\rm out}$ should be proportional to the complex conjugation of the radiated field in (a), given by the distance $r=\sqrt{f^2+(y-y_{\rm f})^2}$ between a point $(y,h)$ on the back of the metalens to the focal spot $(y_{\rm f},h+f)$.}
\label{fig:fig3}
\end{figure}

To ideally focus a windowed (within $|y'|<D_{\rm in}/2$) plane wave $E_x^a(y',z=0)=E_0 e^{ik_y'y'}$ incident from angle $\theta_{\rm in}$ to point ${\bf r}_{\rm f}(\theta_{\rm in})=(y=y_{\rm f}(\theta_{\rm in}), z=h+f)$ on the focal plane, 
%all of the transmitted light should be in phase when reaching ${\bf r}_{\rm f}(\theta_{\rm in})$.
%Therefore, the lens must impart a phase profile that negates (1) the propagation phase from point $(y,z=h)$ on the back surface of the lens to point ${\bf r}_f(\theta_{\rm in})$ and (2) the incident phase profile $k_y' y = (2\pi/\lambda)y\sin\theta_{\rm in}$.
%This requires an ideal phase profile $\phi_{\rm ideal}(y,\theta_{\rm in})$ that varies with the incident angle
the field on the back surface of a metalens should be proportional to the conjugation of the field radiated from a point source at the focal spot to the back surface, as illustrated in Fig.~\ref{fig:fig3}.
Here we consider such ideal transmitted field across the entire back aperture of the lens within $|y| < D_{\rm out}/2$, independent of the incident angle~\cite{note_on_telecentricity}.
The radiated field from a point source in 2D is proportional to $e^{i k r}/\sqrt{r}$, and the distance is $r = \sqrt{f^2 + (y-y_f)^2}$, so the ideal field on the back surface of a metalens is
%\begin{equation}
%\begin{aligned}
%\ \ \ &\phi_{\rm ideal}(x,y,\theta_{\rm in}^x,\theta_{\rm in}^y) \\ =
%&-\frac{2\pi}{\lambda} \left \{\sqrt{f^2+\left (x-x_{\rm f} \right )^2+\left (y-y_{\rm f} \right )^2}+x\sin{\theta_{\rm in}^x}+y\sin{\theta_{\rm in}^y} \right \} \\ &+\psi(\theta_{\rm in}^x,\theta_{\rm in}^y),
%\end{aligned}
%\label{eq:ideal_phase}
%\end{equation}
%where $\psi(\theta_{\rm in}^x,\theta_{\rm in}^y)$ is an angle-dependent constant with no influence on the focusing. The position of the focal spot $\left [ x_{\rm f}(\theta_{\rm in}^x),y_{\rm f}(\theta_{\rm in}^y)) \right ]$ is determined by the incident angle. For each $(\theta_{\rm in}^x,\theta_{\rm in}^y)$, the ideal phase distribution to produce a spherical wavefront varies. This strong angular dependence makes it difficult to achieve diffraction-limited focusing over a wide angular range using a single phase profile and leads to a limited FOV~\cite{aieta2013aberrations,liang2019high}.
%For simplicity, we consider 2D systems, and the ideal phase profile for metalenses under oblique incidence becomes
\begin{equation}
E_x^a(y,z=h)=\begin{cases}
A(\theta_{\rm in})\frac{e^{i\phi_{\rm out}^{\rm ideal}(y,\theta_{\rm in})}}{[f^2+(y-y_{\rm f})^2]^{{1}/{4}}} & \text{for } |y|<\frac{D_{\rm out}}{2} \\
0 & \text{otherwise}
\end{cases},
\label{eq:Ez_at_h}
\end{equation}
where $A(\theta_{\rm in})$ is a constant amplitude, and the ideal phase distribution on the back of the metalens is~\cite{kalvach2016aberration,lalanne2017metalenses,shalaginov2020single}
\begin{equation}
\phi_{\rm out}^{\rm ideal}(y,\theta_{\rm in}) =
\psi(\theta_{\rm in})-\frac{2\pi}{\lambda}\sqrt{f^2+\left [y-y_{\rm f}(\theta_{\rm in}) \right ]^2}.
\label{eq:phase_2D}
\end{equation}
A global phase does not affect focusing, so we include a spatially-constant (but can be angle-dependent) phase function $\psi(\theta_{\rm in})$.
For the focal spot position, we consider $y_{\rm f}(\theta_{\rm in})=f\tan \theta_{\rm in}$, such that the chief ray going through the lens center remains straight.
%But other functions can be used since image distortions can be corrected by software.
A lens system that realizes this angle-dependent phase shift profile $\Delta \phi_{\rm ideal}(y,\theta_{\rm in})=\phi_{\rm out}^{\rm ideal}(y,\theta_{\rm in})-\phi_{\rm in}(y,\theta_{\rm in})$ within the desired $|\theta_{\rm in}| < {\rm FOV}/2$ will achieve diffraction-limited focusing with no aberration, where $\phi_{\rm in}(y,\theta_{\rm in})=(\omega/c)\sin{\theta_{\rm in}}y$ is the phase profile of the incident light.
%Here we consider the whole back surface of the lens to be illuminated, independent of the incident angle~\cite{note_on_telecentricity}.
%Here we are interested in the thinnest-possible lens systems, and for a sufficiently thin lens, the transmitted wavefront will have a negligible angle-dependent lateral displacement compared to the incident wavefront. Therefore, we consider the intensity profile on the back surface of the lens to be uniform across diameter $D_{\rm out}$, independent of the incident angle~\cite{note_on_telecentricity}.
%We can do a discrete Fourier transform on the input side to go from $t(y,\theta_{\rm in})$ to $t(y,y')$, with which we can obtain the lateral spreading $\Delta W(y')$ of interest.

We project the ideal output field in Eq.~\eqref{eq:Ez_at_h} onto a set of flux-orthogonal windowed plane-wave basis to get the angular transmission matrix $t(k_y,k_y')$, as
\begin{equation}
    t_{ba} = \sqrt{\frac{k_z^b}{D_{\rm out}}} %\frac{1}{A\sqrt{D_{\rm in}D_{\rm out}}}\sqrt{\frac{k_z^b}{k_z^a}}
    \int_{-\frac{D_{\rm out}}{2}}^{\frac{D_{\rm out}}{2}} E_x^a(y,z=h)e^{-ik_y^by}dy,
    \label{eq:t_ba}
\end{equation}
where $k_y^a = a(2\pi/D_{\rm in})$ with $a\in \mathbb{Z}$ and  $|k_y^a|<(\omega/c)\sin{(\rm FOV/2)}$, $k_y^b = b(2\pi/D_{\rm out})$ with $b\in \mathbb{Z}$ and $|k_y^b|<\omega/c$, and $(k_y^a)^2+(k_z^a)^2=(k_y^b)^2+(k_z^b)^2=(\omega/c)^2$.
%and $A_a$ is a normalization constant (which can be determined from flux conservation, $\sum_b |t_{ba}|^2=1$, for an ideal lens with unity transmission).
%This integration can be approximated and efficiently computed by a discrete Fourier transform.
The spatial transmission matrix $t(y,y')$ is then given by 
\begin{equation}
    t(y,y') = \frac{1}{\sqrt{D_{\rm in}D_{\rm out}}}\sum_b\sum_a \sqrt{\frac{k_z^a}{k_z^b}}e^{ik_y^by}t_{ba} e^{-ik_y^ay'},
    \label{eq:t_spa}
\end{equation}
where $|y|<D_{\rm out}/2$ and $|y'|<D_{\rm in}/2$.
Detailed derivations and implementations of Eqs.~\eqref{eq:t_ba}--\eqref{eq:t_spa} are given in Supplementary Sec.~2.
From $t(y,y')$, we obtain the lateral spreading $\Delta W(y')$.

\subsection{Thickness bound}

\begin{figure}[tb]
\centering
\includegraphics[width=0.48\textwidth]{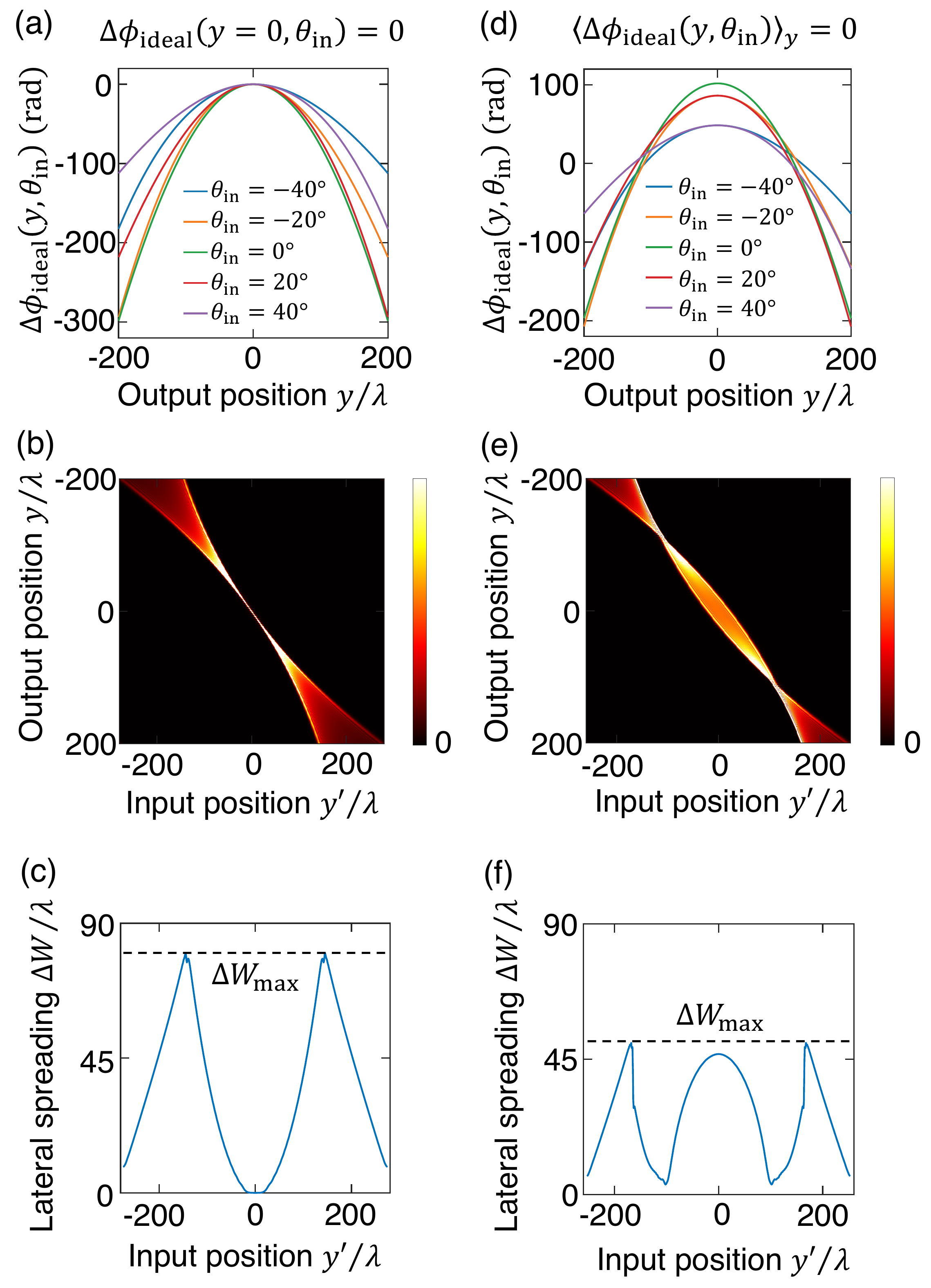}
\caption{Angle-dependent phase shift and lateral spreading of an ideal large-FOV lens.
%the maximal lateral spreading $\Delta W_{\rm max}$ for aberration-free large-FOV lenses. 
(a--c) The incident-angle-dependent phase-shift profiles, spatial transmission matrix $\left |t(y,y') \right |^2$, and lateral spreading $\Delta W(y')$ respectively for an ideal large-FOV lens with the global phase $\psi(\theta_{\rm in})$ chosen such that $\Delta \phi_{\rm ideal}(y=0,\theta_{\rm in})=0$;
%=\frac{2\pi}{\lambda}\sqrt{f^2+y_{\rm f}(\theta_{\rm in})^2}$
this choice minimizes the angle dependence of the phase shift at $y=0$, which minimizes $\Delta W(y'=0)$.
(d--f) Corresponding plots with $\psi(\theta_{\rm in})= \psi_0(\theta_{\rm in})$ in Eq.~\eqref{eq:psi_0}, chosen such that $\langle \Delta \phi_{\rm ideal}(y,\theta_{\rm in}) \rangle_y = 0$ which minimizes $\Delta W_{\rm max}$ and therefore minimizes the thickness bound. Lens parameters: diameter $D_{\rm out}=400\lambda$, ${\rm NA}=0.45$, ${\rm FOV}=80^{\circ}$, with $y_{\rm f}(\theta_{\rm in})=f\tan \theta_{\rm in}$.}
\label{fig:fig4}
\end{figure}

Figure~\ref{fig:fig4}(a--c) plots $\Delta \phi_{\rm ideal}(y,\theta_{\rm in})$, the corresponding transmission matrix in spatial basis, and $\Delta W(y')$ for a lens with output diameter $D_{\rm out}=400\lambda$, ${\rm NA} = \sin(\arctan(D_{\rm out}/(2f)))=0.45$~\cite{note_on_NA}, ${\rm FOV}=80^{\circ}$. 
%Note that we choose $D_{\rm in}>D_{\rm out}$ in the following calculations to remove the periodic replication in the spatial transmission matrix caused by the incompleteness of the truncated plane-wave basis (see Fig.~S3 of Supplement 1). 
Here, the global phase $\psi(\theta_{\rm in})=\frac{2\pi}{\lambda}\sqrt{f^2+y_{\rm f}(\theta_{\rm in})^2}$ is chosen such that $\Delta \phi_{\rm ideal}(y=0,\theta_{\rm in})=0$. 
Note that unlike in Fig.~\ref{fig:fig2}(b), here $\Delta W(y')$ depends strongly on the position $y'$.
%By construction, this $\Delta \phi_{\rm ideal}(y,\theta_{\rm in})$ is angle independent at $y=0$.
An input focused at $y'=0$ is a superposition of plane waves with different angles that constructively interfere at $y'=0$, and since the phase shift $\Delta \phi_{\rm ideal}(y=0,\theta_{\rm in})=0$ is angle-independent there, the transmitted plane waves at different angles still interfere constructively at the output $y=0$, with no lateral spreading, so $\Delta W(y'=0) \approx 0$.
This is analogous to a transform-limited pulse (where its phase is aligned across different frequencies to yield the smallest possible pulse duration) propagating through a dispersion-free medium such that the outgoing pulse is still transform limited with no stretching, but applied to the space-angle Fourier pair instead of the time-frequency Fourier pair. However, away from the lens center, the phase shift $\Delta \phi_{\rm ideal}(y \neq 0,\theta_{\rm in})$ exhibits strong angle dependence as shown in Fig.~\ref{fig:fig4}(a), resulting in significant lateral spreading as shown in Fig.~\ref{fig:fig4}(b--c); this is analogous to pulses that propagate through a medium with strong dispersion and get significantly stretched. %, which will require larger device thickness following Eq.~\eqref{eq:h_and_dW}.

In the above example, $\Delta W_{\rm max} \equiv \max_{y'} \Delta W(y') \approx 80 \lambda$.
Through Eq.~\eqref{eq:h_and_dW}, we can then conclude that such a lens must be at least $80 \lambda$ thick, regardless of how the lens is designed.
This $80 \lambda$ is the axial distance light must propagate in order to accumulate the desired angle-dependent phase shift and the associated lateral spreading.
Recall that $\Delta W$ is also a measure of nonlocality, so the unavoidable lateral spreading here indicates that aberration-free wide-FOV lenses must be nonlocal.
%Indeed, angle-dependent responses are generally nonlocal.

\begin{figure*}[htbp]
\centering
\includegraphics[width=0.95\textwidth]{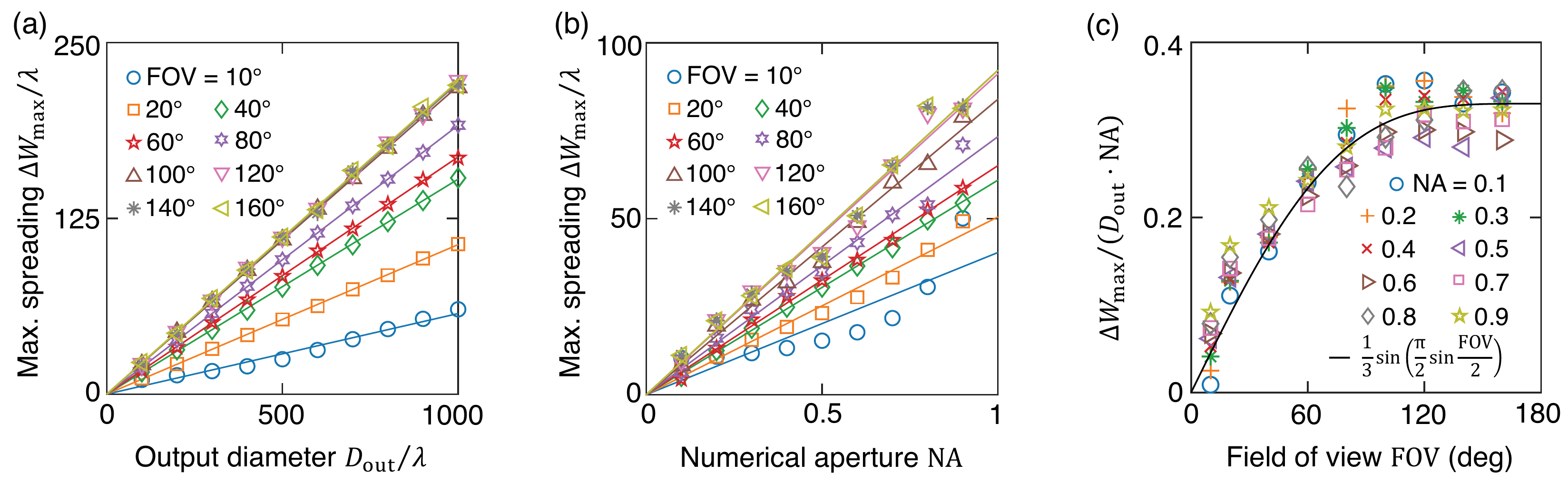}
\caption{Dependence of the optimized maximal lateral spreading $\Delta W_{\rm max}$ on the parameters of an aberration-free wide-FOV lens. (a) $\Delta W_{\rm max}$ as a function of the output diameter $D_{\rm out}$ when ${\rm NA}=0.7$. (b) $\Delta W_{\rm max}$ as a function of the numerical aperture NA when $D_{\rm out}=300\lambda$. %Symbols with different colors represent lateral spreading under different FOV.
Symbols are the maximal spreading of such lenses, and solid lines are linear fits. (c) $\Delta W_{\rm max}/(D_{\rm out} \cdot {\rm NA})$ as a function of the FOV. Black solid line is Eq.~\eqref{eq:DW_max}. % $0.31\sin (\frac{\pi}{2} \sin \frac{\rm FOV}{2})$.
}
\label{fig:fig5}
\end{figure*}

This example uses one particular global phase function $\psi(\theta_{\rm in})=\frac{2\pi}{\lambda}\sqrt{f^2+y_{\rm f}(\theta_{\rm in})^2}$.
Different $\psi(\theta_{\rm in})$ lead to different phase shifts $\Delta \phi_{\rm ideal}(y,\theta_{\rm in})=\phi_{\rm out}^{\rm ideal}(y,\theta_{\rm in})-\phi_{\rm in}(y,\theta_{\rm in})$, with different $\Delta W_{\rm max}$ and different minimal thickness.
Since $\psi(\theta_{\rm in})$ does not affect the focusing quality, we can further lower the thickness bound by optimizing over $\psi(\theta_{\rm in})$ as follows.

%To find the smallest possible thickness $h$ through Eq.~\eqref{eq:h_and_dW}, we want to find the smallest possible $\Delta W_{\rm max}$. 
%This requires minimizing the angle dependence of the phase across the whole surface.
%Therefore, for the thickness bound, we need to find the minimal $\Delta W_{\rm max}$ among possible aberration-free lens configurations.
%Since the global phase function $\psi(\theta_{\rm in})$ does not affect the focusing quality, we use it as free parameters to minimize $\Delta W_{\rm max}$.

\subsection{Minimization of maximal spreading}

%We start with minimization with respect to the global phase $\psi(\theta_{\rm in})$.
%while keeping $y_{\rm f}(\theta_{\rm in})=f\tan \theta_{\rm in}$.
To minimize $\Delta W_{\rm max}$ and the resulting thickness bound, we search for the function $\psi(\theta_{\rm in})$ that minimizes the maximal phase-shift difference among all possible pairs of incident angles across the whole surface,
\begin{equation}
    \mathop{{\rm argmin}}\limits_{\psi(\theta_{\rm in})}\ \max_{y,\theta_{\rm in}^i,\theta_{\rm in}^j} |\Delta \phi_{\rm ideal}(y,\theta_{\rm in}^i; \psi)-\Delta \phi_{\rm ideal}(y,\theta_{\rm in}^j; \psi)|^2,
\label{eq:psi_phase_var}
\end{equation}
where $|y|<D_{\rm out}/2$ and $|\theta_{\rm in}^{i,j}|<{\rm FOV}/2$.

A sensible choice is $\psi(\theta_{\rm in})= \psi_0(\theta_{\rm in})$ with
\begin{equation}
\psi_0(\theta_{\rm in})= \frac{2\pi}{\lambda} \left \langle\sqrt{f^2+\left [y-y_{\rm f}(\theta_{\rm in}) \right ]^2}+y\sin{\theta_{\rm in}} \right \rangle_y
\label{eq:psi_0}
\end{equation}
where $\langle \cdots \rangle_y$ denotes averaging over $y$ within $|y|<D_{\rm out}/2$. With this choice, the phase profiles at different incident angles are all centered around the same $y$-averaged phase, namely $\langle \Delta \phi_{\rm ideal}(y,\theta_{\rm in}) \rangle_y = 0$ for all $\theta_{\rm in}$, so the worst-case variation with respect to $\theta_{\rm in}$ is reduced.
Figure~\ref{fig:fig4}(d--f) shows the resulting phase profile, spatial transmission matrix, and $\Delta W(y')$ with this $\psi=\psi_0$. Indeed, we observe $\Delta W_{\rm max}$ to lower from $80\lambda$ to $50\lambda$ compared to the choice of $\Delta \phi_{\rm ideal}(y=0,\theta_{\rm in})=0$ in Fig.~\ref{fig:fig4}(c).

Eq.~\eqref{eq:psi_phase_var} is a convex problem~\cite{boyd2004convex}, 
%because $\phi_{\rm ideal}(y,\theta_{\rm in}; \psi, y_{\rm f})$ is a linear function of $\psi$,
so its global minimum can be found with established algorithms.
We use the CVX package~\cite{cvx,gb08} to perform this convex optimization.
Section 3 and Fig.~S5 of Supplementary Materials show that the $\psi_0(\theta_{\rm in})$ in Eq.~\eqref{eq:psi_0} is very close to the global optimum of Eq.~\eqref{eq:psi_phase_var}, and the two give almost identical $\Delta W_{\rm max}$.
%As another verification, we also directly minimize $\Delta W_{\rm max}$ using gradient-based nonlinear optimization~\cite{NLopt,svanberg2002class} of an equivalent epigraph formulation of the minimax problem, and it yields similar $\Delta W_{\rm max}$ and $\psi(\theta_{\rm in})$ as shown in Fig.~S2. 
Therefore, in the following we adopt the $\psi_0(\theta_{\rm in})$ in Eq.~\eqref{eq:psi_0} to obtain the smallest-possible thickness bound.

One can potentially also vary the focal spot position $y_{\rm f}(\theta_{\rm in})$ to further minimize $\Delta W_{\rm max}$, since image distortions can be corrected by software. After optimizing over $y_{\rm f}$, % ({\it i.e.}, $y_{\rm f}=f\sin{\theta_{\rm in}}$ and random numbers), 
we find that $y_{\rm f}(\theta_{\rm in})=f\tan{\theta_{\rm in}}$ already provides close-to-minimal $\Delta W_{\rm max}$.
%For the focal spot position, $y_{\rm f}(\theta_{\rm in})=f\tan \theta_{\rm in}$ is commonly assumed, but other functions can be used since image distortions can be corrected by software.
%So we also explore if $y_{\rm f}(\theta_{\rm in})$ can be chosen to further reduce $\Delta W_{\rm max}$.
%Figures~S3-4 and Sec.~2B of Supplement 1 show that $y_{\rm f}(\theta_{\rm in})=f\tan \theta_{\rm in}$ already gives a small phase variation when NA is less than approximately 0.5, while a different $y_{\rm f}(\theta_{\rm in})$ can be used to slightly reduce $\Delta W_{\rm max}$ for lenses with NA larger than 0.5.

%An example of the ideal phase profile with optimized $\psi(\theta_{\rm in})$ and $y_{\rm f}(\theta_{\rm in})$ is shown in Fig.~\ref{fig:fig3}(c), exhibiting a smaller phase variation compared to that of the conventional case in Fig.~\ref{fig:fig3}(a).
%Figure~\ref{fig:fig3}(d) shows the spatial transmission matrix obtained from $\phi_{\rm ideal}(y,\theta_{\rm in})$ plotted in Fig.~\ref{fig:fig3}(c). The maximal lateral spreading reduces significantly as expected.

\subsection{Dependence on lens parameters}

The above procedure can be applied to any wide-FOV lens. For example, we now know that the lens considered in Figure~\ref{fig:fig4} must be at least $50\lambda$ thick regardless of its design.
It is helpful to also know how such a minimal thickness depends on the lens parameters, so we carry out a systematic study here.
%Next we map out how the optimized $\Delta W_{\rm max}$ depends on the lens parameters. %, using transmission matrices from $\phi_{\rm ideal}(y,\theta_{\rm in})$ with $\psi(\theta_{\rm in})=\psi_0(\theta_{\rm in})$ and the numerically optimized $y_{\rm f}(\theta_{\rm in})$. 

%For small and intermediate FOV, each vertical slice $|t(y,y')|^2$ of the ideal lens is close to a rectangular function, for which the IPR in \eqref{eq:W_out} well captures the output width.
Supplementary Video~1 shows how the ideal transmission matrix in both bases evolve as the FOV increases.
While increasing the FOV only adds more columns to the angular transmission matrix, doing so increases the variation of the phase shift with respect to the incident angle ({\it i.e.}, increases the angular diversity), which changes the spatial transmission matrix and increases the lateral spreading ({\it i.e.}, increases nonlocality).
We also observe that the output profiles in $|t(y,y')|^2$ develop two strong peaks at the edges as the FOV increases.
The IPR in Eq.~\eqref{eq:W_out} is better suited for functions that are unimodal or close to rectangular.
Therefore, when FOV $\ge 100^{\circ}$, we use the full width at half maximum (FWHM) instead to quantify $W_{\rm out}$; Figure~S7 of the Supplementary Materials shows that the FWHM equals IPR for small FOV but is a better measure of the output width for large FOV.

\begin{table*}[t]
\newcommand{\tabincell}[2]{\begin{tabular}{@{}#1@{}}#2\end{tabular}}
\centering
\caption{\bf Metalenses with diffraction-limited focusing over a wide FOV$^{\rm a}$}
\begin{threeparttable}
\begingroup
\setlength{\tabcolsep}{5.5pt} % increase column spacing
\renewcommand{\arraystretch}{1.1} % increase row spacing
\begin{tabular}{cccccccccc}
\hline
  & Method &  \tabincell{c}{Exp./\\Sim.} & \tabincell{c}{Output diameter\\$D_{\rm out}$($D_{\rm out}^{\rm eff}$)} & \tabincell{c}{Numerical\\aperture} & \tabincell{c}{FOV\\(air)} & \tabincell{c}{Strehl\\ratio} & \tabincell{c}{Total\\thickness} & \tabincell{c}{Thickness\\bound} \\
\hline
Arbabi \emph{et al.}~\cite{arbabi2016miniature} & Doublet & 3D Exp. & (800 $\upmu$m) & 0.49 & $60^{\circ}$ & $\gtrsim 0.9$ & 1 mm & 92 $\upmu$m \\
Groever \emph{et al.}~\cite{groever2017meta} & Doublet & 3D Exp. & (313 $\upmu$m) & 0.44 & $50^{\circ}$ & $\gtrsim 0.8$ & 500 $\upmu$m & 30 $\upmu$m \\
He \emph{et al.}~\cite{he2019polarization} & Doublet & 3D Sim. & (400 $\upmu$m) & 0.47 & $60^{\circ}$ & - & 500 $\upmu$m & 44 $\upmu$m \\
Li \emph{et al.}~\cite{li2021super} & Doublet & 3D Sim. & (20 $\upmu$m) & 0.45 & $50^{\circ}$ & $\gtrsim 0.5$ & 31.2 $\upmu$m & 1.8 $\upmu$m \\
Tang \emph{et al.}~\cite{tang2020achromatic} & Doublet & 3D Sim. & (30 $\upmu$m) & 0.35 & $40^{\circ}$ & - & 21.2 $\upmu$m & 1.8 $\upmu$m \\
Kim \emph{et al.}~\cite{kim2020doublet} & Doublet & 3D Sim. & (300 $\upmu$m) & 0.38 & $60^{\circ}$ & - & 500 $\upmu$m & 27 $\upmu$m \\
Huang \emph{et al.}~\cite{huang2021achromatic} & Doublet & 3D Sim. & (5 $\upmu$m) & 0.60 & $60^{\circ}$ & - & 6.6 $\upmu$m & 0.7 $\upmu$m \\
Engelberg \emph{et al.}~\cite{engelberg2020near} & Aperture & 3D Exp. & (1.35 mm) & 0.20 & $30^{\circ}$ & - & 3.36 mm & 0.03 mm \\
\tabincell{c}{Shalaginov \emph{et al.}~\cite{shalaginov2020single} \\ Shalaginov \emph{et al.}~\cite{shalaginov2020single}} & \tabincell{c}{Aperture \\ Aperture} & \tabincell{c}{3D Exp. \\ 3D Sim.} & \tabincell{c}{(1 mm) \\ (1 mm)} & \tabincell{c}{0.24 \\ 0.20} & \tabincell{c}{$\sim 180^{\circ}$ \\ $\sim 180^{\circ}$} & \tabincell{c}{$\gtrsim 0.8$ \\ $\gtrsim 0.8$} & \tabincell{c}{2 mm \\ 3.9 mm} & \tabincell{c}{0.08 mm \\ 0.07 mm} \\
Fan \emph{et al.}~\cite{fan2020ultrawide} & Aperture & 3D Sim. & (20 $\upmu$m) & 0.25 & $170^{\circ}$ & $\gtrsim 0.8$ & $38.6\ \upmu$m & 1.7 $\upmu$m \\
Zhang \emph{et al.}~\cite{zhang2021extreme} & Aperture & 3D Exp. & (1 mm) & 0.11 & $\sim 180^{\circ}$ & - & 5.44 mm & 0.04 mm \\
Yang \emph{et al.}~\cite{yang2021design} & Aperture & 3D Sim. & (100 $\upmu$m) & 0.18 & $\sim 180^{\circ}$ & $\sim 0.64$ & 200 $\upmu$m & 6 $\upmu$m \\
Lin \emph{et al.}~\cite{lin2018topology} & Multi-layer & 2D Sim. & $23 \lambda$ & 0.35 & $40^{\circ}$ & - & $1.5 \lambda$ & $1.4 \lambda$ \\

\tabincell{c}{Lin \emph{et al.}~\cite{lin2021computational} \\ Lin \emph{et al.}~\cite{lin2021computational} \\ Lin \emph{et al.}~\cite{lin2021computational}} & \tabincell{c}{Multi-layer \\ Multi-layer \\ Multi-layer} & \tabincell{c}{2D Sim. \\ 2D Sim. \\ 3D Sim.} & \tabincell{c}{$50 \lambda$ \\ $125 \lambda$ \\ $50 \lambda$} & \tabincell{c}{0.24 \\ 0.70 \\ 0.12} & \tabincell{c}{$60^{\circ}$ \\ $80^{\circ}$ \\ $16^{\circ}$} & \tabincell{c}{$\gtrsim 0.8$ \\ $\gtrsim 0.8$ \\ $\gtrsim 0.8$} & \tabincell{c}{$12 \lambda$ \\ $24 \lambda$ \\ $12 \lambda$} & \tabincell{c}{$2.8 \lambda$ \\ $25 \lambda$ \\ $0.4 \lambda$} \\
\hline
\end{tabular}
\endgroup
 \begin{tablenotes}
        \footnotesize
        \item [a] We note that the thickness bound here is directly from Eq.~\eqref{eq:thickness}, which is an approximate expression and is obtained for 2D systems but suffices as an estimation.
        Refs.~\cite{arbabi2016miniature,groever2017meta,he2019polarization,li2021super,tang2020achromatic,kim2020doublet,huang2021achromatic,engelberg2020near,shalaginov2020single,fan2020ultrawide,zhang2021extreme,yang2021design} adopt a telecentric configuration where each incident angle fills an effective diameter $D_{\rm out}^{\rm eff}$ within the output aperture, which we use in place of $D_{\rm out}$ when evaluating their thickness bounds.
        Some works also correct the chromatic aberration: at 473 nm and 532 nm in Ref.~\cite{tang2020achromatic}, at 445 nm, 532 nm and 660 nm in Ref.~\cite{kim2020doublet}, from 470 nm to 650 nm in Ref.~\cite{huang2021achromatic}, and from 1 to 1.2 $\upmu$m in Ref.~\cite{yang2021design}. Ref.~\cite{lin2018topology} achieves diffraction-limited focusing for 7 angles within the FOV. Ref.~\cite{lin2021computational} achieves diffraction-limited focusing for 19, 7 and 9 angles within the FOV and also corrects the chromatic aberration for 10, 4, and 5 frequencies within a 23\% spectral bandwidth from up to down. 
\end{tablenotes}
\end{threeparttable}
\label{tab:tab1}
\end{table*}

Next, we quantify the dependence on all lens parameters.
Figure~\ref{fig:fig5} plots the optimized maximal lateral spreading $\Delta W_{\rm max}$ as a function of the output diameter $D_{\rm out}$, NA and the FOV. As shown in Fig.~\ref{fig:fig5}(a), $\Delta W_{\rm max}$ grows linearly with $D_{\rm out}$ for different FOV. Figure~\ref{fig:fig5}(b) further shows that $\Delta W_{\rm max}$ also grows approximately linearly with the numerical aperture NA.
Figure~\ref{fig:fig5}(a,b) fix ${\rm NA}=0.7$ and $D_{\rm out}=300\lambda$ respectively, while similar dependencies are observed for other lens parameters (Figs.~S8--9 of Supplementary Materials). Dividing by $D_{\rm out}$ and NA, we obtain how $\Delta W_{\rm max}$ depends on the FOV, shown in Fig.~\ref{fig:fig5}(c). 
The angular range is governed by $\sin({\rm FOV}/2)$, but the functional dependence of $\Delta W_{\rm max}$ on the FOV is not simply $\sin({\rm FOV}/2)$; empirically, we find the function $\frac{1}{3}\sin (\frac{\pi}{2} \sin \frac{\rm FOV}{2})$ to provide a reasonable fit
for the FOV dependence. These dependencies can be summarized as
\begin{equation}
\Delta W_{\rm max}\approx (\frac{1}{3}{\rm NA})D_{\rm out}\sin \left (\frac{\pi}{2} \sin \frac{\rm FOV}{2} \right ).
\label{eq:DW_max}
\end{equation}
Eq.~\eqref{eq:h_and_dW} and Eq.~\eqref{eq:DW_max} then tell us approximately how the thickness bound varies with the lens parameters,
\begin{equation}
h\gtrsim (\frac{1}{3}{\rm NA})D_{\rm out}\sin \left (\frac{\pi}{2} \sin \frac{\rm FOV}{2} \right ).
\label{eq:thickness}
\end{equation}
This result makes intuitive sense, since increasing the NA, aperture size, and/or FOV will all lead to an increased phase-shift variation, which leads to the increased minimal thickness.

%\subsection{Thickness bound}

Any aberration-free wide-FOV lens system must have a transmission matrix described in Sec.~\ref{sec:DW}, so the above bound applies to any such system regardless of how the system is designed (barring unlikely broad-angle resonant effects).
%including cascaded metasurfaces, diffractive lenses, bulk metamaterials and thick volumetric structures.
This result shows that to achieve large FOV with a wide output aperture, a single layer of subwavelength-thick metasurface is fundamentally not sufficient. Meanwhile, it also reveals room to make existing designs more compact, as we discuss below.
%Composite structures with a sufficient total thickness, such as those shown in Fig.~\ref{fig:fig1}(c), are a must.

While the results above are obtained for 2D systems, we expect qualitatively similar results in 3D (possibly only with a slightly different prefactor) since the relation between angular diversity and lateral spreading and the relation between lateral spreading and thickness are both generic.
Note that we use FOV to denote the range of incident angles from air.
Equation~\eqref{eq:thickness} continues to hold in the presence of substrates, with the Snell's law $\sin \frac{\rm FOV}{2}=n_{\rm in}\sin \frac{\rm FOV_{in}}{2}$ for ${\rm FOV}_{\rm in}$ in a substrate with refractive index $n_{\rm in}$, since we have shown in Fig.~S1 that Eq.~\eqref{eq:h_and_dW} holds in the presence of a substrate and since the ideal transmission matrix in Sec.~\ref{sec:DW} is the same with or without a substrate.

%\subsection{Comparison with existing methods}

%Several implementations have been proved successful for the achievement of diffraction-limited focusing over a broad angular range, such as adding an aperture stop, using a metasurface doublet or triplet, and designing multi-layer metasurfaces. We then compare our thickness bound in Eq.~\eqref{eq:thickness} to various
Table~\ref{tab:tab1} lists diffraction-limited wide-FOV metalens systems reported in the literature. % and compares their total thickness with the bound in Eq.~\eqref{eq:thickness}. 
All of them have total thickness consistent with Eq.~\eqref{eq:thickness}. A few inverse-designed multi-layer structures~\cite{lin2018topology,lin2021computational} have thickness close to the bound, suggesting that the bound is tight. Note that the second design in Ref.~\cite{lin2021computational} has a slightly smaller thickness ($24\lambda$) than the bound ($25\lambda$), likely because it only optimizes for diffraction-limited focusing at a discrete set of angles. Existing metalenses based on doublets or aperture stops are substantially thicker than the bound, which is sensible since those systems have ample amount of free spaces not used for structural design.
%a further analysis on the required total thickness of the aperture stop approach is given in Supplementary Sec.~6.

Here we consider ideal aberration-free focusing for all incident angles within the FOV. Relaxing some of these conditions can relax the thickness bound; for example, if diffraction-limited focusing is not necessary, the quadratic phase profile~\cite{pu2017nanoapertures,martins2020metalenses,lassalle2021imaging} can eliminate the angle dependence of the phase profile.
Meanwhile, %Eq.~\eqref{eq:thickness} considers monochromatic waves.
%and the focal spot function $y_{\rm f}(\theta_{\rm in})=f\tan \theta_{\rm in}$.
achromatic wide-FOV lenses~\cite{shrestha2019multi,tang2020achromatic,kim2020doublet,huang2021achromatic,lin2021computational,yang2021design} will be subject to additional constraints beyond nonlocality~\cite{shastri2022bandwidth}.
%or when other focal spot functions are used (such as Fourier lenses where $f\sin \theta_{\rm in}$ is desirable, or when the output is stretched to emulate a virtual focal length~\cite{lin2021computational}),
%the minimal device thickness will be larger, which can be the subject of a future work.

\section{Discussion}

%Let us summarize the origin of this thickness bound: ideal wide-FOV lenses must exhibit an incident-angle-dependent response [Eq.~\eqref{eq:phase_2D}], the Fourier transform duality between momentum and position %[Eq.~\eqref{eq:t_change_basis}]
%translates that angular variation to a spatial spreading that characterizes nonlocality (Fig.~\ref{fig:fig4}), %[Eq.~\eqref{eq:lateral_spreading}]
Due to the Fourier-transform duality between space and momentum, any multi-channel system with an angle-dependent response will necessarily require nonlocality and spatial spreading (exemplified in Fig.~\ref{fig:fig4} and analogous to a pulse propagating through a dispersive medium under time-frequency duality), 
which is tied to the device thickness through Eq.~\eqref{eq:h_and_dW}.
This relationship is not limited to wide-FOV lenses and establishes the intrinsic link between angular diversity and spatial footprint suggested in the introduction.

For example, one can readily use this approach to establish thickness bounds for other types of nonlocal metasurfaces such as retroreflectors~\cite{arbabi2017planar} and photovoltaic concentrators~\cite{price2015wide,shameli2018absorption,lin2018topology,roques2022towards} where a wide angular range is also desirable. Note that concentrators are additionally subject to efficiency bounds arising from passivity and/or reciprocity~\cite{zhang2019scattering}. 

These results can guide the design of future nonlocal metasurfaces, providing realistic targets for device dimensions.
While multi-layer metasurfaces that reach Eq.~\eqref{eq:thickness} have not been experimentally realized yet, there are several realistic routes. A stacked triple-layer metalens has been reported~\cite{shrestha2019multi}. Multi-layer structures have been realized with two-photon polymerization~\cite{christiansen2020fullwave,roques2022towards,2022_Roberts_cleo},
or repeated deposition and patterning of 2D layers~\cite{2011_Sherwood-Droz_OE,zhou2018multilayer,mansouree2020multifunctional,camayd2020multifunctional}.
Volumetric nanostructures may also be realized with deposition onto shrinking scaffolds~\cite{oran20183d}.
Additionally, multi-level diffractive lenses can readily have thickness above 10 $\mu$m~\cite{2019_Meem_PNAS,meem2021imaging}.

Fundamental bounds like this are valuable as metasurface research evolves beyond single-layer local designs, as better control of light is achieved over wider ranges of angles, and with the continued push toward ultra-compact photonic devices.
Future work can investigate designs incorporating broad-angle resonant responses.
We also note that the transmission-matrix approach is versatile and can be used to establish other types of bounds beyond the device footprint.

%\begin{backmatter}
%\bmsection{Funding} Content in the funding section will be generated entirely from details submitted to Prism. Authors may add placeholder text in the manuscript to assess length, but any text added to this section in the manuscript will be replaced during production and will display official funder names along with any grant numbers provided. If additional details about a funder are required, they may be added to the Acknowledgments, even if this duplicates information in the funding section. See the example below in Acknowledgements.

\section*{Acknowledgements} 
We thank O.~D.~Miller, H.-C.~Lin, and X.~Gao for helpful discussions.
This work is supported by the National Science Foundation CAREER award (ECCS-2146021) and the Sony Research Award Program.
\bf{Author contributions:} \rm S. Li performed the calculations, optimizations, and data analysis; C.W.H. proposed the initial idea and supervised research; both contributed to designing the study, discussing the results, and preparing the manuscript.
\bf{Competing interests:} \rm The authors declare no competing interests.
\bf{Data and materials availability:} \rm All data needed to evaluate the conclusions in this study are presented in the paper and in the supplementary materials.

%\bmsection{Supplemental document}
%See Supplement 1 for supporting content. 

%\end{backmatter}

%\newpage
% Bibliography
\bibliography{maintext}

\begin{thebibliography}{10}
\bibitem{note_on_telecentricity}
\bibinfo{note}{The angular distribution of the output depends on the incident
  angle, so the lens is not telecentric.}
\bibitem{note_on_NA}
\bibinfo{note}{We define NA based on normal incidence.}
\end{thebibliography}
\bibliographystyle{naturemag}

% Full bibliography added automatically for Optics Letters submissions; the following line will simply be ignored if submitting to other journals.
% Note that this extra page will not count against page length
%\bibliographyfullrefs{sample}

\end{document}

\maketitle

\section{Lateral spreading versus thickness}
\label{sec:DW_substrate}

\begin{figure}[htbp]
\centering
\includegraphics[width=0.4\linewidth]{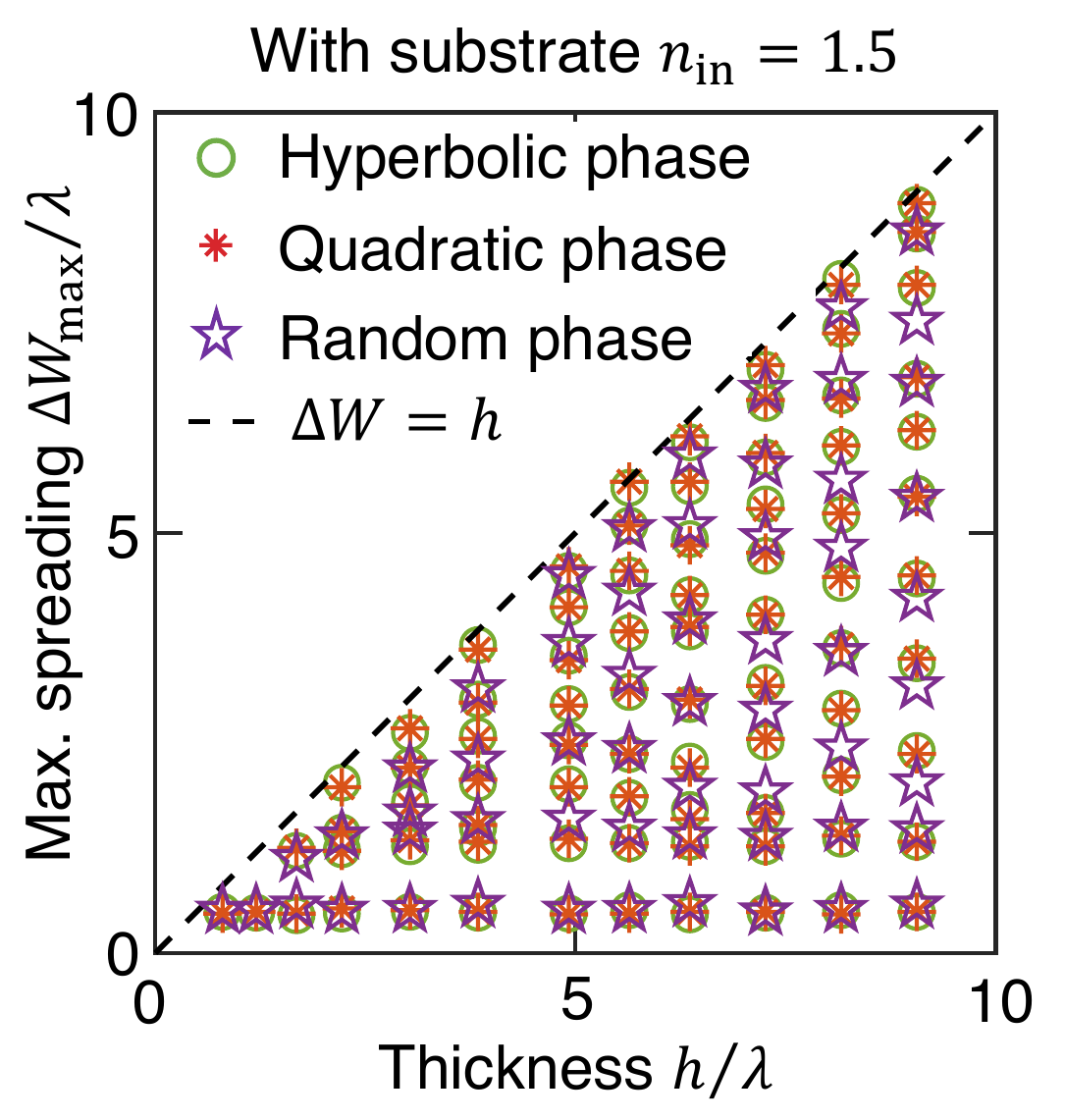}
\caption{Maximal lateral spreading $\Delta W_{\rm max}$ computed from the spatial transmission matrix, for different metasurfaces of varying thicknesses on a substrate with refractive index $n_{\rm in}=1.5$ at FOV = $180^{\circ}$.}
\label{fig:DW_substrate}
\end{figure}

\begin{figure}[htbp]
\centering
\includegraphics[width=0.98\linewidth]{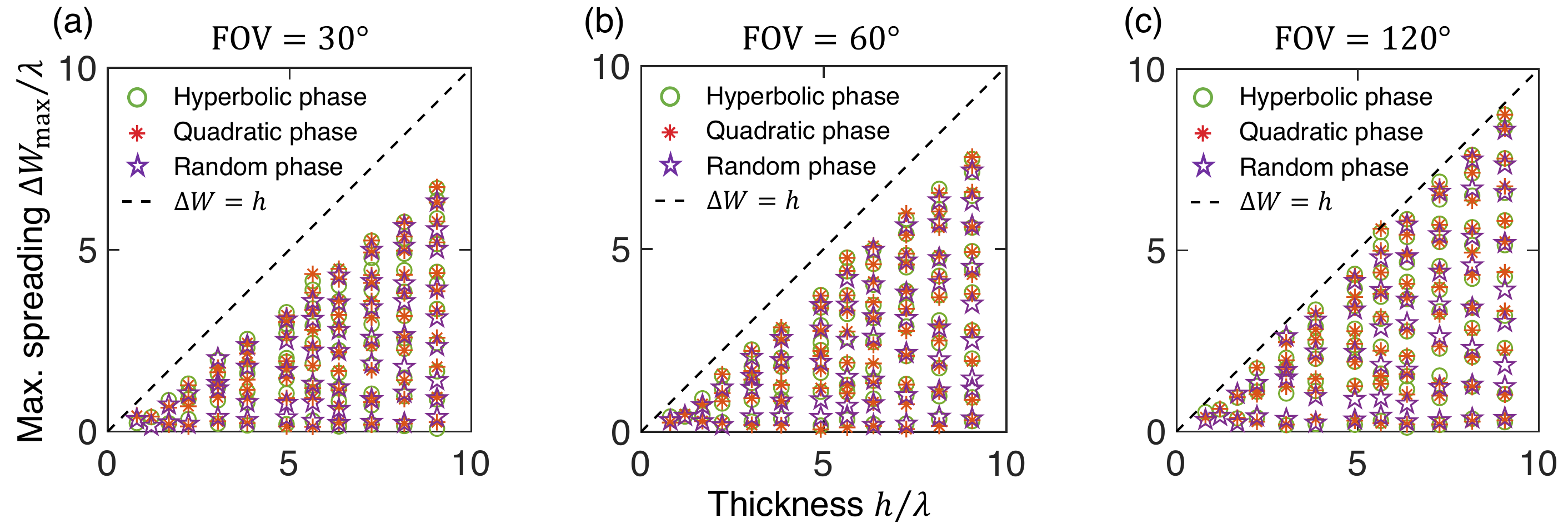}
\caption{Maximal lateral spreading $\Delta W_{\rm max}$ computed from the spatial transmission matrix, for different metasurfaces of varying thicknesses with no substrate ($n_{\rm in}=1$) at smaller values of FOV.}
\label{fig:DW_air_FOV}
\end{figure}

Figure~2(d) of the main text plots the maximal lateral spreading $\Delta W_{\rm max}$ as a function of the thickness $h$ for various metasurfaces with no substrate ($n_{\rm in}=1$) at FOV = $180^{\circ}$.
Figure~\ref{fig:DW_substrate} shows $\Delta W_{\rm max}$ of the same metasurfaces placed on a substrate with refractive index $n_{\rm in}=1.5$ at FOV = $180^{\circ}$.
We observe the same relation here: $\Delta W_{\rm max} < h$. This relation also holds when different FOV is considered, as shown in Fig.~\ref{fig:DW_air_FOV}.

\section{Transmission matrix of ideal wide-FOV metalenses}
\label{sec:flux_norm_TM}
\subsection{Basis of transverse channels}
\label{sec:basis}

\begin{figure}[htbp]
\centering
\includegraphics[width=.38\linewidth]{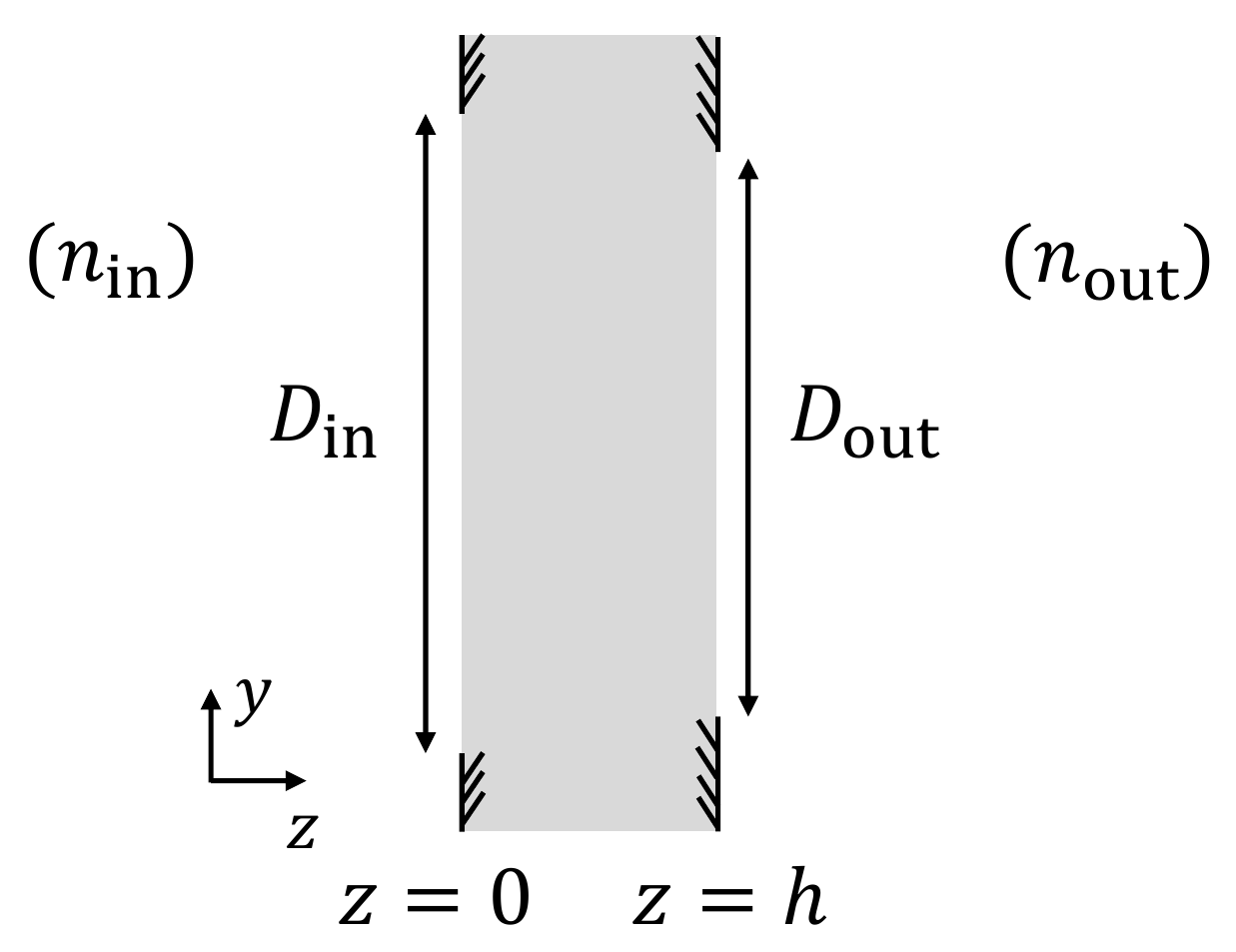}
\caption{Schematic of the system: light incident from a medium with refractive index $n_{\rm in}$ onto an entrance aperture with diameter $D_{\rm in}$, propagates through the structure with thickness $h$, and exits through an output aperture with diameter $D_{\rm out}$ into a medium with refractive index $n_{\rm out}$ .}
\label{fig:2d_system}
\end{figure}

Consider the transverse magnetic (TM) fields of a 2D system, with ${\bf E}=E_x(y,z)\hat{x}$, where the refractive index is $n_{\rm in}$ and $n_{\rm out}$ on the incident ($z<0$) and transmitted ($z>h$) sides respectively, as shown in Fig.~\ref{fig:2d_system}. The Poynting flux for a monochromatic wave at $\omega$ is
\begin{equation}
    {\bf S}=\frac{1}{2}{\rm Re}[{\bf E}^*\times {\bf H}]=\frac{1}{2\omega \mu_0}{\rm Im}\left(0,E_x^*\frac{\partial E_x}{\partial y},E_x^*\frac{\partial E_x}{\partial z}\right).
\label{eq:poynting}
\end{equation}
%For an ideal system with 100\% transmission efficiency, the incoming flux at $z=0$ should be equal to the transmitted flux at $z=h$, which means that ${\rm Im}\int _{-\infty}^{\infty}dy\ E_x^*\frac{\partial E_x}{\partial z}|_{z=0}={\rm Im}\int _{-\infty}^{\infty}dy\ E_x^*\frac{\partial E_x}{\partial z}|_{z=h}$.

We expand the incident field near the input aperture, $E_x^{\rm in}(y,z \lesssim 0)$, in a basis of truncated plane waves:
\begin{equation}
   E_x^{\rm in}(y,z\lesssim 0)=\sum _{a}v_af_a(y,z),
\label{eq:Ex_in}
\end{equation}
where
\begin{equation}
   f_a(y,z)=\begin{cases}
   \frac{1}{\sqrt{D_{\rm in}}}\frac{1}{\sqrt{k_z^a}}e^{i(k_y^ay+k_z^az)}  & \text{for } |y|<\frac{D_{\rm in}}{2}\\
   0  & \text{otherwise}
   \end{cases},
\label{eq:in_basis}
\end{equation}
with $(k_y^a)^2+(k_z^a)^2=(n_{\rm in}\omega/c)^2$ for channel $a$, and $\{k_y^a\}$ are real-valued. The $\frac{1}{\sqrt{D_{\rm in}}}\frac{1}{\sqrt{k_z^a}}$ prefactors are used to normalize the longitudinal flux. The incident flux in $z$ direction is then proportional to
\begin{equation}
%\begin{aligned}
{\rm Im}\int _{-\infty}^{\infty}dy \left . (E_x^{\rm in})^*\frac{\partial E_x^{\rm in}}{\partial z} \right|_{z=0}
%= &\frac{1}{D_{\rm in}}{\rm Im}\int _{-\frac{D_{\rm in}}{2}}^{\frac{D_{\rm in}}{2}}dy\ \left[\sum_{a}v_a^*\frac{e^{-ik_y^ay}}{\sqrt{k_z^a}}\right]\left[\sum_{b}v_bik_z^b\frac{e^{ik_y^by}}{\sqrt{k_z^b}}\right] \\ =
=\sum_{a}|v_a|^2+{\rm Re}\sum_{a}\sum_{b\neq a}v_a^*v_b\sqrt{\frac{k_z^b}{k_z^a}}\frac{\sin{[(k_y^b-k_y^a)D_{\rm in}/2]}}{2(k_y^b-k_y^a)D_{\rm in}},
%\end{aligned}
\label{eq:in_flux}
\end{equation}
where the summation only includes propagating channels (with real-valued $k_z^a$) as the evanescent ones do not carry flux.

Based on the Nyquist-Shannon sampling theorem~\cite{landau1967sampling}, a discrete sampling of the momentum space with spacing $2\pi/D_{\rm in}$ is sufficient to represent a band-limited function with bandwidth $D_{\rm in}$ in real space.
Therefore, we consider input wave numbers
\begin{equation}
    \{k_y^a\} = \left\{ a\frac{2\pi}{D_{\rm in}} \text{ such that } a\in \mathbb{Z} \text{ and } |k_y^a|<\frac{\omega}{c} \sin{\left(\rm \frac{FOV}{2}\right)} \right\},
\label{eq:ky_in}
\end{equation}
limited to within the FOV of interest. Note that $\sin \frac{\rm FOV}{2}=n_{\rm in}\sin \frac{\rm FOV_{in}}{2}$.
With this choice, the different basis functions of Eq.~(\ref{eq:in_basis}) are orthogonal, and the cross terms in Eq.~(\ref{eq:in_flux}) drop out.
%So we will only consider the space of incident waves specified by Eqs.~(\ref{eq:Ex_in})(\ref{eq:in_basis})(\ref{eq:ky_in}).

Similarly, the transmitted field $E_x(y,z\gtrsim h)$ near the output aperture can be expanded in a flux-orthogonal basis of truncated plane waves
\begin{equation}
   E_x(y,z\gtrsim h)=\sum _{b}u_bg_b(y,z),
\label{eq:Ez_out}
\end{equation}
where
\begin{equation}
   g_b(y,z)=\begin{cases}
   \frac{1}{\sqrt{D_{\rm out}}}\frac{1}{\sqrt{k_z^b}}e^{i[k_y^by+k_z^b(z-h)]}  & \text{for } |y|<\frac{D_{\rm out}}{2}\\
   0  & \text{otherwise}
   \end{cases},
\label{eq:out_basis}
\end{equation}
with
\begin{equation}
    \{k_y^b\} = \left\{ b\frac{2\pi}{D_{\rm out}} \text{ such that } b\in \mathbb{Z} \text{ and } |k_y^b|<\frac{\omega}{c}n_{\rm out} \right\}.
\label{eq:ky_out}
\end{equation}
and $(k_y^b)^2+(k_z^b)^2=(n_{\rm out}\omega/c)^2$. Note that the transmitted field can also have evanescent components, but here we only consider contributions from the propagating ones.

%Imposing the condition of 100\% transmission efficiency to the power-orthogonal basis in Eqs.~(\ref{eq:in_basis})(\ref{eq:out_basis}), there should be
%\begin{equation}
    %\sum_{a}|v_a|^2 = \sum_{b} |u_b|^2.
%\label{eq:effi_1}
%\end{equation}
%Define the angular transmission matrix $t(k_y,k_y^{\prime})$ as $u=t(k_y,k_y^{\prime})v$, then it should satisfy
%\begin{equation}
    %t^{\dagger}(k_y,k_y^{\prime})t(k_y,k_y^{\prime})=\mathbbm{1}.
%\label{eq:effi_2}
%\end{equation}

%Above, we consider the full transmission matrix with $|k_y^a|<n_{\rm in}\omega/c$ and $|k_y^b|<n_{\rm out}\omega/c$. We can also restrict the angular transmission matrix $t(k_y,k_y^{\prime})$ to incident angles within the FOV of interest by imposing $|k_y^a|<(n_{\rm in}\omega/c)\sin{(\rm FOV_{in}/2)}$ with $\rm FOV_{in}$ defined as the angular range in the input medium [{\it i.e.}, $n_{\rm in}\rm \sin{(FOV_{in}/2)}=\sin{(FOV/2)}$], and all of the remaining results still apply. 

\subsection{Transmission matrix in angular basis}
\label{sec:ang_TM}
Consider incident wave from a fixed angle
%a general 2D metalens with $D_{\rm in}\neq D_{\rm out}$, whose incident and transmitted fields are described by
\begin{equation}
E_x^a(y',z=0)= f_a(y',z=0),
%\begin{cases}
%Ae^{ik_{\rm y}^{\prime}y'} & \text{for } |y'|<D_{\rm in}/2 \\
%0 & \text{otherwise}
%\end{cases},
\label{eq:Ex_at_0}
\end{equation}
so $v_{a'} = \delta_{a a'}$. Following section IIA of the main text, the transmitted field for an ideal lens is
\begin{equation}
E_x^a(y,z=h)=\begin{cases}
A(\theta_{\rm in})\frac{e^{i\phi_{\rm out}^{\rm ideal}(y,\theta_{\rm in})}}{[f^2+(y-y_{\rm f})^2]^{\frac{1}{4}}} & \text{for } |y|<D_{\rm out}/2 \\
0 & \text{otherwise}
\end{cases},
\label{eq:Ex_at_h}
\end{equation}
which we can project onto the preceding basis $\{g_b\}$ to get
%\begin{equation}
%    v_a = \sqrt{\frac{k_z^a}{D_{\rm in}}} \int_{-\frac{D_{\rm in}}{2}}^{\frac{D_{\rm in}}{2}}E_x(y',z=0)e^{-ik_y^ay'}dy',
%\label{eq:v_a}
%\end{equation}
\begin{equation}
    u_b = \sqrt{\frac{k_z^b}{D_{\rm out}}} \int_{-\frac{D_{\rm out}}{2}}^{\frac{D_{\rm out}}{2}}E_x^a(y,z=h)e^{-ik_y^by}dy.
\label{eq:u_b}
\end{equation}
Since $u_b=\sum_{a'}t_{ba'}v_{a'} = t_{ba}$, this gives the $a$-th column of the transmission matrix in angular basis. To summarize, we have
%the angular transmission matrix $t_{ba}=t(k_y^b,k_y^a)$, with $k_y^a$ and $k_y^b$ representing the input and output channels respectively, is defined as $u_b=\sum_{a}t_{ba}v_a$, it can be obtained for a wide-FOV metalens as
\begin{equation}
\begin{aligned}
    &t_{ba} = \sqrt{\frac{k_z^b}{D_{\rm out}}} %\frac{1}{A}\frac{1}{\sqrt{D_{\rm in}D_{\rm out}}}\sqrt{\frac{k_z^b}{k_z^a}}
    \int_{-\frac{D_{\rm out}}{2}}^{\frac{D_{\rm out}}{2}} E_x^a(y,z=h)e^{-ik_y^by}dy, \\
    \text{with } &(k_y^a)^2+(k_z^a)^2=(\frac{\omega}{c}n_{\rm in})^2,\ k_y^a=a\frac{2\pi}{D_{\rm in}},\ a\in \mathbb{Z},\ |k_y^a|<\frac{\omega}{c}\sin{\left(\rm \frac{FOV}{2}\right)} \\
    &(k_y^b)^2+(k_z^b)^2=(\frac{\omega}{c}n_{\rm out})^2,\ k_y^b=b\frac{2\pi}{D_{\rm out}},\ b\in \mathbb{Z},\ |k_y^b|<\frac{\omega}{c}n_{\rm out}.
\end{aligned}
\label{eq:t_ba}
\end{equation}
%where
%\begin{equation}
    %\Phi(y,k_y^a)=\begin{cases}
    %\frac{e^{i\phi_{\rm out}^{\rm ideal}(y,\theta_{\rm in}; %\psi)}}{[f^2+(y-y_{\rm f})^2]^{\frac{1}{4}}} & \text{for } |y|<D_{\rm out}/2 \\
    %0 & \text{otherwise}
    %\end{cases}.
%\label{eq:Psi}
%\end{equation}
The normalization constant $A(\theta_{\rm in})$ can be determined from flux conservation, $\sum_b |t_{ba}|^2=1$, for an ideal lens with unity transmission.

%Note that $\{k_y^a\}$ and $\{k_y^b\}$ in Eq.~(\ref{eq:t_ba}) are discretized, but the integration over $y$ is continuous. From the information point of view, a continuous spatial sampling is redundant, and the discrete sampling of space with spacing $\frac{\lambda/n_{\rm out}}{2}$ for the output is sufficient according to the Nyquist-Shannon sampling theorem. But in order to maintain the accuracy of the integration, a finer spacing is needed for the output $y$. Similar attentions also need to be paid on the calculation of inverse participation ratio (IPR) in Eq.~(6) of the main text, which also involves the integration over $y$. Finally, we choose the output spacing $\Delta y=\frac{\lambda/n_{\rm out}}{4}$ to balance the accuracy and speed.
To evaluate the transmission matrix in practice, we approximate the continuous integration over $y$ in Eq.~(\ref{eq:t_ba}) and in the evaluation of the inverse participation ratio (IPR) [Eq.~(7) of the main text] with a discrete summation with spacing $\Delta y$.
To determine the resolution to use, we use $\Delta y=\lambda/4$ and $\Delta y=\lambda/20$ to evaluate the maximal lateral spreading $\Delta W_{\rm max}$ for $D_{\rm out}=200\lambda$ or $900\lambda$, NA = 0.2 or 0.8, FOV = $40^{\circ}$ or $160^{\circ}$. 
The relative difference of $\Delta W_{\rm max}$ between the two choices of $\Delta y$ is smaller than 5\% in all cases, so we use $\Delta y=\lambda/4$ in the following.
%, thus $\Delta y=\frac{\lambda/n_{\rm out}}{4}$ provides sufficient accuracy for the discretization of both Eq.~(\ref{eq:t_ba}) and Eq.~(6) of the main text.

Eq.~(\ref{eq:t_ba}) approximated with a discrete summation can be evaluated efficiently using fast Fourier transform (FFT),
\begin{equation}
    t_{ba} \approx \Delta y \sqrt{\frac{k_z^b}{D_{\rm out}}} % \frac{1}{A}\frac{\Delta y}{\sqrt{D_{\rm in}D_{\rm out}}}\sqrt{\frac{k_z^b}{k_z^a}}
    e^{-ik_y^b(-\frac{D_{\rm out}}{2}+\frac{\Delta y}{2})}\sum_{n=0}^{N-1}E_x^a(y_n,z=h)e^{-i\frac{2\pi}{N}b n},
\label{eq:t_ba_fft}
\end{equation}
where $N\equiv D_{\rm out}/\Delta y\in \mathbb{Z}$ is the number of segments the interval $y\in [-D_{\rm out}/2,D_{\rm out}/2]$ is discretized into, and $\{y_n\equiv -\frac{D_{\rm out}}{2}+\frac{\Delta y}{2}+n\Delta y\}$ are the centers of those segments. This produces $N$ rows of the transmission matrix with $0 \le b \le N-1$; we cyclically rearrange the output index $b$ and keep the propagating channels within $|k_y^b|<n_{\rm out}\omega/c$.

%Note that the amplitude $A(\theta_{\rm in})$ of the output field $E_x(y,z=h)$ at varied incident angles is chosen by $\sum_{b}|t_{ba}|^2=1$ for all the $a$ involved, which conserves energy at different incident angles.
%To achieve 100\% transmission efficiency ({\it i.e.}, Eq.~(\ref{eq:effi_2})), we should project the angular transmission efficiency $t(k_y,k_y^{\prime})$ onto the unitary space through a singular value decomposition
%\begin{equation}
    %t(k_y,k_y^{\prime})=U\Sigma V^{\dagger},
%\label{eq:svd}
%\end{equation}
%where $U$ and $V$ are unitary matrices; $\Sigma$ is a rectangular diagonal matrix with non-negative singular values $\{\sigma\}$ on the diagonal. The condition number $\kappa$ defined as the ratio between the maximal and minimal singular values, {\it i.e.} $\kappa=\sigma_{\rm max}/\sigma_{\rm min}$, can be used to evaluate the unitarity of $t(k_y,k_y^{\prime})$. By reducing $D_{\rm in}$ from $D_{\rm out}$, we can expect a reasonably small $\kappa$ and close-to-unitary $t(k_y,k_y^{\prime})$ [add ref]. Then by replacing all of the singular values $\{\sigma\}$ by 1, {\it i.e.} $t(k_y,k_y^{\prime})=UV^{\dagger}$, we can get the angular transmission matrix of a wide-FOV metalens with 100\% transmission matrix without degrading the diffraction-limited performance (See Sec.~7 for the comparison of Strehl ratio).

\subsection{Transmission matrix in spatial basis}
\label{sec:spa_TM}

%After obtaining angular transmission matrices for ideal wide-FOV metalenses, we can then do a change of basis from $t(k_y,k_y^{\prime})$ to $t(y,y^{\prime})$.

We define the transmission matrix in spatial basis, $t(y,y^{\prime})$, by
\begin{equation}
    E_x(y,z=h) \triangleq \int_{-\infty}^{\infty} t(y,y')E_x^{\rm in}(y',z=0)\ dy',
\label{eq:spa_TM_1}
\end{equation}
where only propagating waves within the input and output apertures are kept in the incident field $E_x^{\rm in}(y',z=0)$ and in the transmitted field $E_x(y,z=h)$.
%For a given $E_x(y',z=0)$ and $E_x(y,z=h)$, we remove their components beyond $|y'|<D_{\rm in}/2$ and $|y|<D_{\rm out}/2$ and evanescent waves, respectively.

From Eqs.~(\ref{eq:Ez_out},\ref{eq:out_basis}) with
$u_b=\sum_{a}t_{ba}v_{a}$ and
$v_a = \sqrt{\frac{k_z^a}{D_{\rm in}}} \int_{-\frac{D_{\rm in}}{2}}^{\frac{D_{\rm in}}{2}}E_x^{\rm in}(y',z=0)e^{-ik_y^ay'}dy'$, we have
\begin{equation}
    E_x(y,z=h) = \frac{1}{\sqrt{D_{\rm in}D_{\rm out}}}\sum_b\sum_a \sqrt{\frac{k_z^a}{k_z^b}} e^{ik_y^by} t_{ba}\int_{-\frac{D_{\rm in}}{2}}^{\frac{D_{\rm in}}{2}}E_x^{\rm in}(y',z=0)e^{-ik_y^ay'}\ dy'
\label{eq:spa_TM_2}
\end{equation}
when $|y|<D_{\rm out}/2$. Comparing Eqs.~(\ref{eq:spa_TM_1})(\ref{eq:spa_TM_2}), we obtain the spatial transmission matrix
\begin{equation}
    t(y,y') = \begin{cases} \frac{1}{\sqrt{D_{\rm in}D_{\rm out}}}\sum_b\sum_a \sqrt{\frac{k_z^a}{k_z^b}} e^{ik_y^by} t_{ba} e^{-ik_y^ay'} & \text{when } |y|<D_{\rm out}/2 \text{ and } |y'|<D_{\rm in}/2 \\ 0 & \text{otherwise}
    \end{cases}.
\label{eq:spa_TM_continuous}
\end{equation}

In Eq.~(\ref{eq:spa_TM_continuous}), the spatial coordinates $y$ and $y'$ are both continuous variables, which is redundant from the information point of view. Since the transmission matrix has a bandwidth of $|k_y|<(\omega/c)n_{\rm out}$ in the output and a bandwidth of $|k_y'|<(\omega/c)\sin{({\rm FOV}/2)}$ in the input, a discrete sampling of space with spacing $\Delta y=\frac{\lambda/n_{\rm out}}{2}$ in the output and spacing $\Delta y'=\frac{\lambda}{2\sin{(\rm FOV/2)}}$ in the input should be sufficient, which is the Nyquist sampling rate. 
But accurate reconstruction would then require the Whittaker--Shannon interpolation.
We skip the interpolation and simply use a finer resolution $\Delta y=\lambda/4$ in the output as mentioned in Sec.~\ref{sec:flux_norm_TM}B. Since no integration is needed over the input $y'$, we sample the input with spacing $\Delta y'=\frac{\lambda}{2\sin{(\rm FOV/2)}}$. 
%After doing the discretization of $y$ and $y'$, Eq.~(\ref{eq:spa_TM_continuous}) becomes a matrix with the same size as the angular transmission matrix.
Eq.~(\ref{eq:spa_TM_continuous}) with discrete sampling can also be evaluated efficiently with FFT and inverse FFT.

\begin{comment}
We can also use FFT to efficiently compute the spatial transmission matrix by
\begin{equation}
    t(y_n,y_{n'}') = \frac{1}{\sqrt{D_{\rm in}D_{\rm out}}}\sum_b\sum_a e^{i\frac{2\pi}{N}(n-1)b}\Tilde{t}_{ba}e^{-i\frac{2\pi}{N'}(n'-1)a},
\label{eq:spa_TM_fft}
\end{equation}
where
\begin{equation}
    \Tilde{t}_{ba} = \begin{cases} e^{i\frac{2\pi}{D_{\rm out}}(-\frac{D_{\rm out}}{2}+\frac{\Delta y}{2})b}\sqrt{\frac{k_z^a}{k_z^b}}t_{ba}e^{-i\frac{2\pi}{D_{\rm in}}(-\frac{D_{\rm in}}{2}+\frac{\Delta y'}{2})a} & \text{when } |k_y^b|<\frac{\omega}{c}n_{\rm out} \text{ and } |k_y^a|<\frac{\omega}{c}n_{\rm in}\sin{(\frac{\rm FOV_{in}}{2})} \\ 0 & \text{otherwise}
    \end{cases},
\label{eq:t_ba_tilde}
\end{equation}
$N\equiv D_{\rm out}/\Delta y\in \mathbb{Z}$ and $N'\equiv D_{\rm in}/\Delta y'\in \mathbb{Z}$ give the number of the discrete segments for the output $y\in [-D_{\rm out},D_{\rm out}/2]$ and input $y'\in [-D_{\rm in}/2,D_{\rm in}/2]$, respectively; $y_n\equiv -\frac{D_{\rm out}}{2}+\frac{\Delta y}{2}+(n-1)\Delta y,\ n=1,2,\cdots,N$ and $y'_{n'}\equiv -\frac{D_{\rm in}}{2}+\frac{\Delta y'}{2}+(n'-1)\Delta y',\ n'=1,2,\cdots,N'$ denote the center of the segments.
\end{comment}

\begin{figure}[htbp]
\centering
\includegraphics[width=1\linewidth]{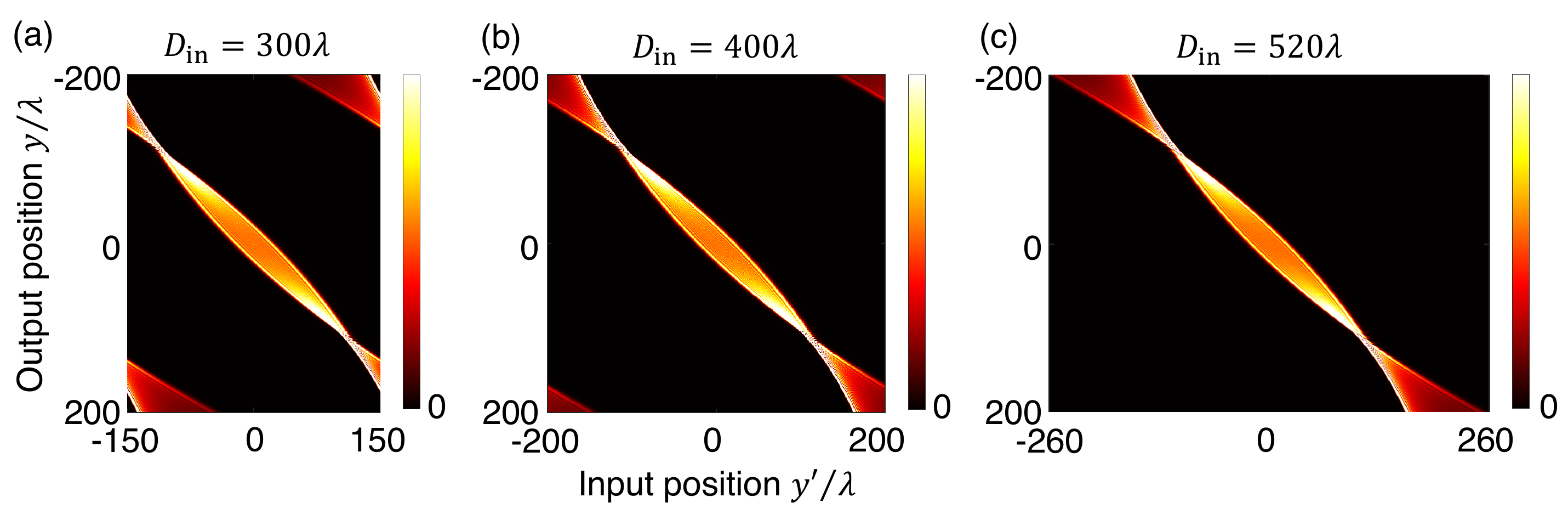}
\caption{Spatial transmission matrix $|t(y,y')|^2$ of an ideal wide-FOV lens with different $D_{\rm in}$. As $D_{\rm in}$ increases, the periodic replication goes away. Lens parameters: diameter $D_{\rm out}=400\lambda$, ${\rm NA}=0.45$, ${\rm FOV}=80^{\circ}$, with $\psi(\theta_{\rm in})=\psi_0(\theta_{\rm in})$ and $y_{\rm f}(\theta_{\rm in})=f\tan \theta_{\rm in}$.}
\label{fig:spa_TM}
\end{figure}

Note that in order to make the transverse channels orthogonal, we imposed in Sec.~\ref{sec:flux_norm_TM}A that $k_y^a$ and $k_y^b$ have spacing $2\pi/D_{\rm in}$ and $2\pi/D_{\rm out}$, which makes $E_x^{\rm in}(y',z=0)$ and $E_x(y,z=h)$ periodic in $y'$ and $y$ with a period of $D_{\rm in}$ and $D_{\rm out}$ respectively. 
Such an artificial periodic boundary leads to periodic replications in the spatial transmission matrix, as shown in Fig.~\ref{fig:spa_TM}, which makes the lateral spreading inaccurate for $y'$ near the two edges. %, mainly due to the periodicity of the input. 
To remove this unphysical periodic replication while maintaining orthogonality of the basis, we choose a $D_{\rm in}$ that is larger than $D_{\rm out}$ (specifically, $D_{\rm in}=5D_{\rm out}$). % and plot the spatial transmission matrix within a smaller range of $D_{\rm in}$ to remove the blackness. This is acceptable because extra $D_{\rm in}$ will not influence the accuracy of $\Delta W_{\rm max}$ as can be seen in Fig.~\ref{fig:spa_TM}.
Such choice does not affect the conclusion of this study since the angular variation of the phase shift (and subsequently $\Delta W_{\rm max}$ and the minimal thickness) depends on $D_{\rm out}$, not $D_{\rm in}$.

\section{Optimization of global phase}
\label{sec:opt_2D}

The function $\psi(\theta_{\rm in})$ in Eq.~(11) of the main text does not affect focusing quality, so we use it as a free parameter to minimize the maximal variation of $\Delta \phi_{\rm ideal}(y,\theta_{\rm in})$ with respect to $\theta_{\rm in}$, which minimizes $\Delta W_{\rm max}$ and the associated thickness bound.

%As shown in Fig.~3 of the main text, conventionally, $\psi$ with respect to $\theta_{\rm in}$ is set to $\psi(\theta_{\rm in})=\frac{2\pi}{\lambda}\sqrt{f^2+y_{\rm f}(\theta_{\rm in})^2}$ to force $\phi_{\rm ideal}(y=0,\theta_{\rm in})=0$. This $\psi(\theta_{\rm in})$ can also be set to 
%\begin{equation}
%\psi(\theta_{\rm in})=\left \langle \frac{2\pi}{\lambda} \left \{\sqrt{f^2+\left [y-y_{\rm f}(\theta_{\rm in}) \right ]^2}+y\sin{\theta_{\rm in}} \right \} \right \rangle_y
%\label{eq:s2}
%\end{equation}
%to compensate the $y$-averaged phase, equivalent to shifting $\phi_{\rm ideal}(y,\theta_{\rm in})$ under varied $\theta_{\rm in}$ to the same level. Intuitively, this $\psi(\theta_{\rm in})$ should be (or at least close to) the optimal choice that can minimize the phase variation of $\phi_{\rm ideal}(y,\theta_{\rm in})$. It is obvious in Fig.~3 of the main text that setting $\psi(\theta_{\rm in})$ to be the $y$-averaged phase can further reduce the phase variation of $\phi_{\rm ideal}(y,\theta_{\rm in})$ and thus reduce the lateral spreading compared to the conventional case.

As described in the main text, a sensible choice is 
\begin{equation}
    \psi(\theta_{\rm in})= \frac{2\pi}{\lambda} \left \langle \sqrt{f^2+\left [y-y_{\rm f}(\theta_{\rm in}) \right ]^2}+y\sin{\theta_{\rm in}} \right \rangle_y \equiv \psi_0(\theta_{\rm in}),
\label{eq:psi}
\end{equation}
where $\langle \cdots \rangle_y$ denotes averaging over $y$ within $|y|<D_{\rm out}/2$.
%With this choice, the phase profile at different incident angles are all centered around the same $y$-averaged phase, namely $\langle \phi_{\rm ideal}(y,\theta_{\rm in}) \rangle_y = 0$ for all $\theta_{\rm in}$, so we expect the worst-case variation with respect to $\theta_{\rm in}$ to be reduced.

To assess the performance of the $\psi_0(\theta_{\rm in})$ above, we carry out a numerical optimization where $\psi(\theta_{\rm in})$ is varied to minimize the maximal phase-shift difference between all possible pairs of incident angles across all positions:
\begin{equation}
    \mathop{{\rm argmin}}\limits_{\psi(\theta_{\rm in})}\  \max_{y,\theta_{\rm in}^i,\theta_{\rm in}^j} |\Delta \phi_{\rm ideal}(y,\theta_{\rm in}^i; \psi)-\Delta \phi_{\rm ideal}(y,\theta_{\rm in}^j; \psi)|^2,
\label{eq:psi_phase_var}
\end{equation}
where $\theta_{\rm in}^i$ and $\theta_{\rm in}^j$ represent different incident angles with $|\theta_{\rm in}^{i,j}|<{\rm FOV}/2$, $|y|<D_{\rm out}/2$, and
\begin{equation}
\begin{aligned}
\Delta \phi_{\rm ideal}(y,\theta_{\rm in};\psi) &=
\phi_{\rm out}^{\rm ideal}(y,\theta_{\rm in})-\phi_{\rm in}(y,\theta_{\rm in}) \\
&=
\psi(\theta_{\rm in})-\frac{2\pi}{\lambda}\left[ \sqrt{f^2+\left [y-y_{\rm f}(\theta_{\rm in}) \right ]^2} + y \sin{\theta_{\rm in}} \right].
\end{aligned}
\end{equation}
\begin{comment}
This minimax problem can be equivalently formulated~\cite{boyd2004convex} by introducing a dummy variable $u$,
\begin{equation}
\begin{aligned}
    &\mathop{{\rm argmin}}\limits_{u,\psi(\theta_{\rm in})}\ u, \\ &{\rm subject\ to}\ u\geqslant |\phi_{\rm ideal}(y,\theta_{\rm in}^i)-\phi_{\rm ideal}(y,\theta_{\rm in}^j)|^2.
\end{aligned}
\label{eq:minimax}
\end{equation}
\end{comment}
This is a convex optimization problem~\cite{boyd2004convex} because $\Delta \phi_{\rm ideal}(y,\theta_{\rm in}; \psi)$ is a linear function of $\psi$. We use CVX~\cite{cvx,gb08}, a package for specifying and solving convex programming problems, to find the global optimum of Eq.~(\ref{eq:psi_phase_var}).

\begin{figure}[tbp]
\centering
\includegraphics[width=.85\linewidth]{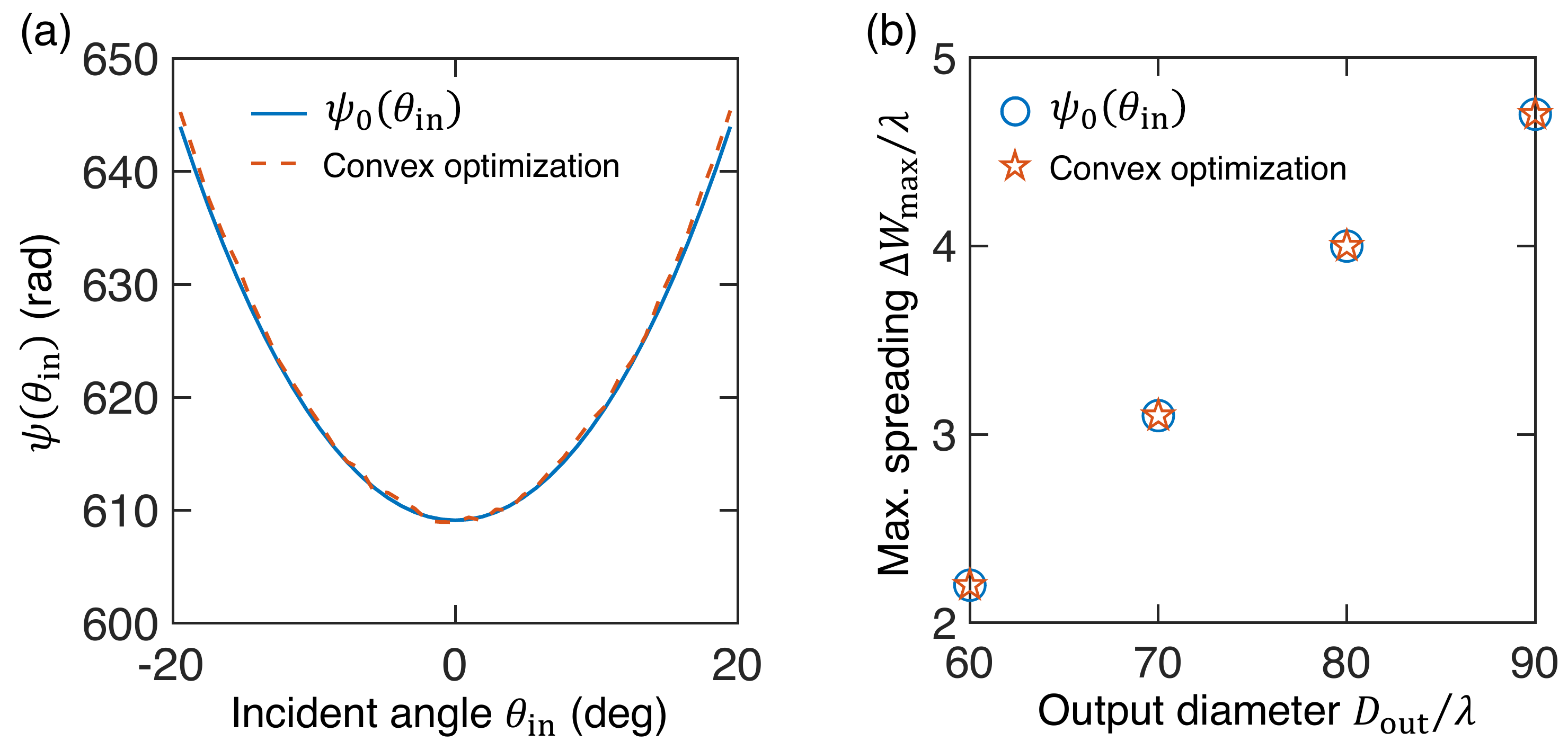}
\caption{Optimization with respect to $\psi(\theta_{\rm in})$. (a) $\psi_0(\theta_{\rm in})$ in Eq.~(\ref{eq:psi}) and the $\psi(\theta_{\rm in})$ from optimizing Eq.~(\ref{eq:psi_phase_var}).
Lens parameters: diameter $D_{\rm out}=60\lambda$, ${\rm NA}=0.3$, ${\rm FOV}=40^{\circ}$, $y_{\rm f}(\theta_{\rm in})=f\tan \theta_{\rm in}$.
(b) The maximal lateral spreading $\Delta W_{\rm max}$ as a function of the output diameter $D_{\rm out}$ after optimizing $\psi(\theta_{\rm in})$.
}
\label{fig:psi_opt}
\end{figure}

The global optimum of $\psi(\theta_{\rm in})$ is in close agreement with the $\psi_0(\theta_{\rm in})$ in Eq.~(\ref{eq:psi}), as shown in Fig.~\ref{fig:psi_opt}(a).
While there are small differences among the two, such differences have no noticeable effect on the resulting maximal lateral spreading $\Delta W_{\rm max}$, as shown in Fig.~\ref{fig:psi_opt}(b).
Therefore, in the following, we use $\psi=\psi_0$ in Eq.~(\ref{eq:psi}) as the global phase.

\section{FOV dependence (animation)}
\label{sec:animation}
\begin{figure}[htbp]
\centering
\includegraphics[width=1\linewidth]{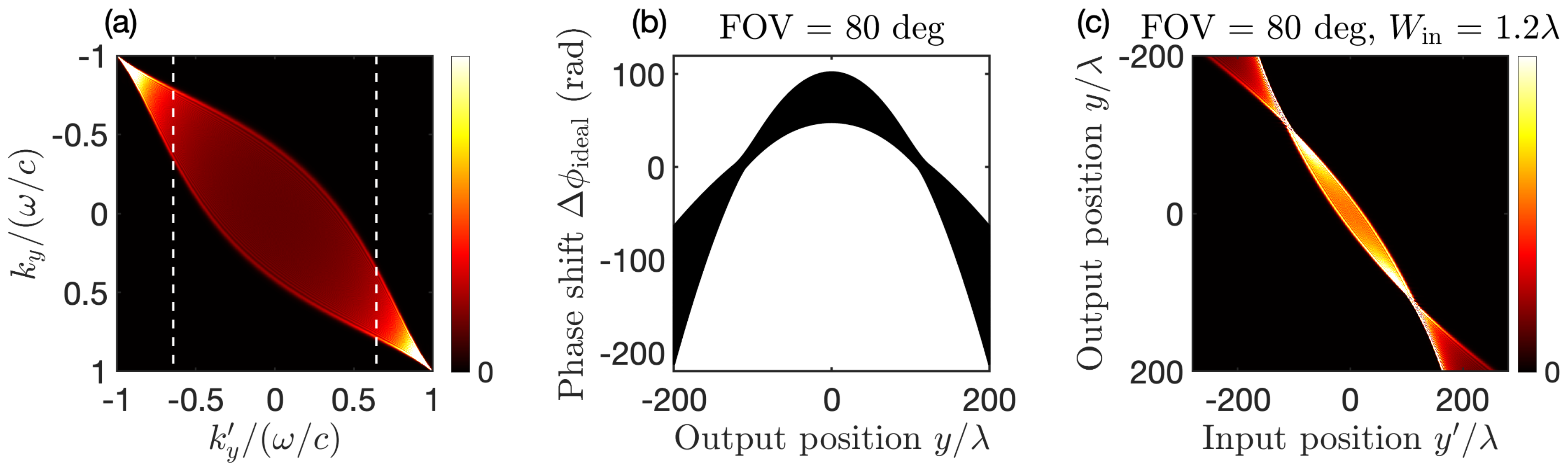}
\caption{One frame of Supplementary Video 1, which shows how the transmission matrix of an ideal lens evolves with the FOV. (a) The angular transmission matrix $|t(k_y,k_y')|^2$; white dashed lines show the boundary given by the FOV. (b) Phase-shift profiles across incident angles within the FOV. (c) The corresponding spatial transmission matrix $\left |t(y,y') \right |^2$ at the given FOV.
Lens parameters: diameter $D_{\rm out}=400\lambda$, ${\rm NA}=0.45$, with $\psi(\theta_{\rm in})=\psi_0(\theta_{\rm in})$ and $y_{\rm f}(\theta_{\rm in})=f\tan \theta_{\rm in}$.}
\label{fig:animation}
\end{figure}

Supplementary Video 1 shows how the ideal transmission matrix (in both bases) and the phase-shift profiles evolve as the FOV increases.
Figure~\ref{fig:animation} provides the animation caption and shows one frame of the animation. Increasing the FOV leads to an increased phase-shift variation among incident angles, which widens the diagonal of the spatial transmission matrix. 

\section{Output profiles and output widths}
\label{sec:output_width}
\begin{figure}[htbp]
\centering
\includegraphics[width=1\linewidth]{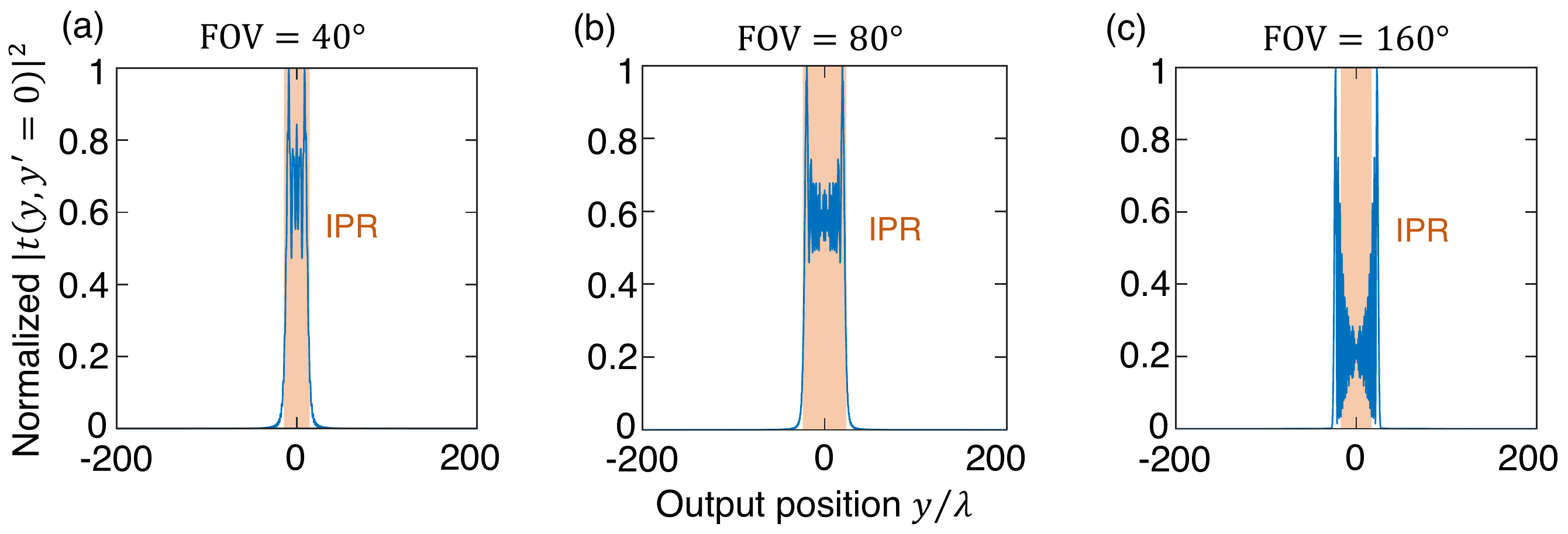}
\caption{Output profiles of the ideal spatial transmission matrix (blue lines), with the widths defined by the inverse participation ratio (IPR) indicated with orange shadings. Lens parameters: $D_{\rm out}=400\lambda$, NA = 0.45, with $\psi(\theta_{\rm in})=\psi_0(\theta_{\rm in})$ and $y_{\rm f}(\theta_{\rm in})=f\tan \theta_{\rm in}$.}
\label{fig:output_width}
\end{figure}

Figure~\ref{fig:output_width} shows the middle column of the spatial transmission matrix for ideal wide-FOV lenses with different FOVs.
With a small FOV, the output profiles are reasonably close to being rectangular, and the inverse participation ratio (IPR) coincides with the full width at half maximum (FWHM).
With a large FOV, the output profiles develop two strong peaks, and the IPR underestimates the output width.

\section{Comprehensive data for dependence on lens parameters}
\label{sec:LS}

\begin{figure}[htbp]
\centering
\includegraphics[width=.98\linewidth]{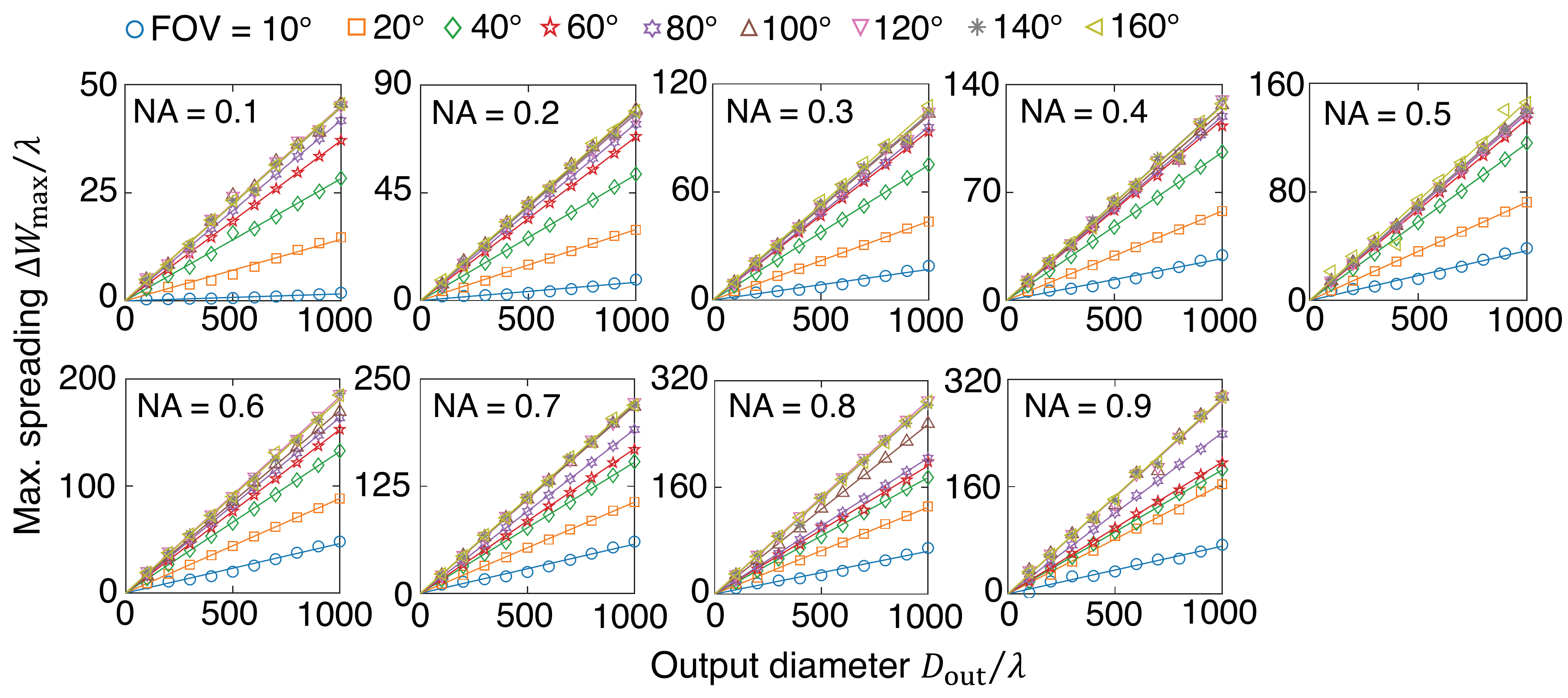}
\caption{Optimized maximal lateral spreading as a function of the output diameter $D_{\rm out}$ for different NA and FOV.}
\label{fig:dW_D}
\end{figure}

\begin{figure}[htbp]
\centering
\includegraphics[width=.96\linewidth]{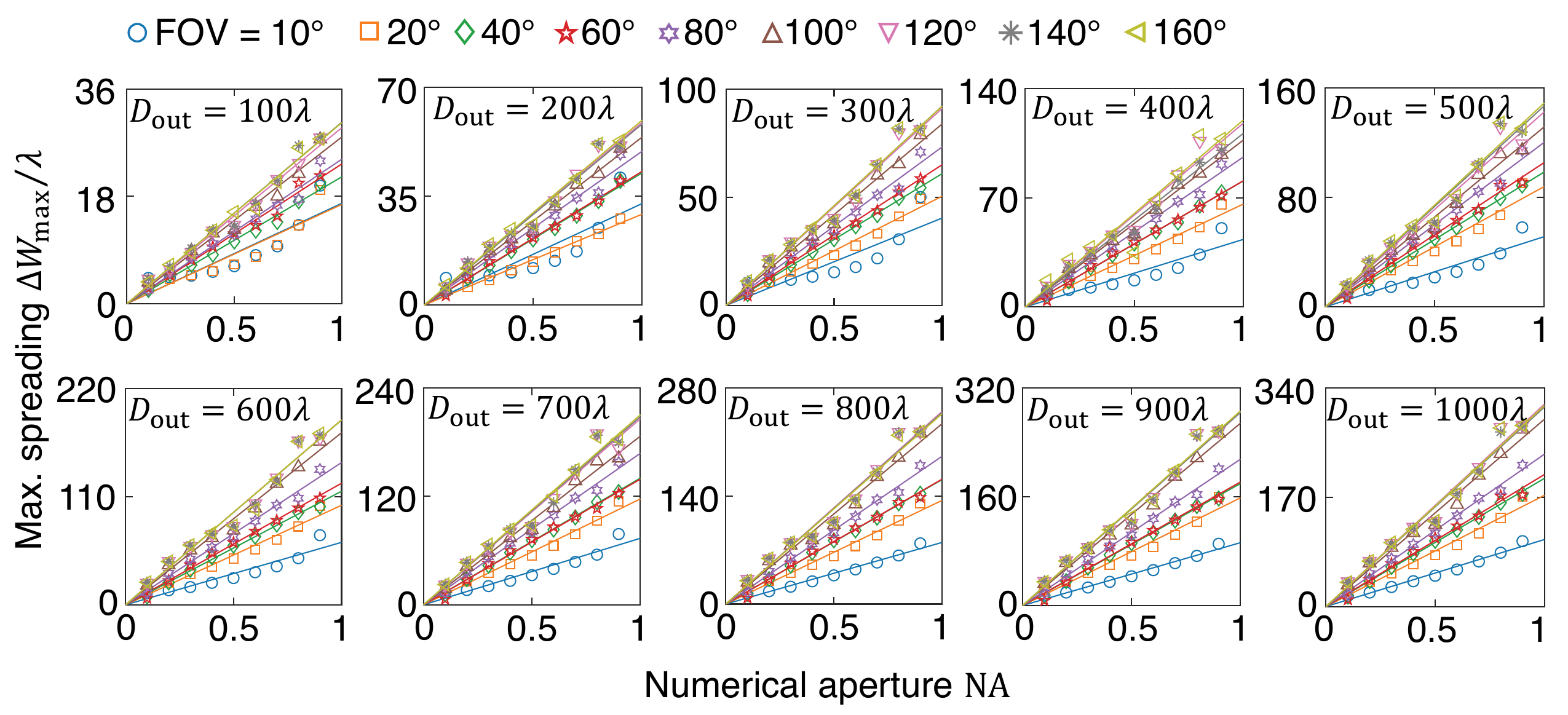}
\caption{Optimized maximal lateral spreading as a function of the numerical aperture NA for different lens diameter $D_{\rm out}$ and FOV.}
\label{fig:dW_NA}
\end{figure}

Figure~5(a,b) of the main text plots the optimized maximal lateral spreading $\Delta W_{\rm max}$ for ${\rm NA}=0.7$ and output diameter $D_{\rm out}=300\lambda$ respectively, with varying FOV.
Figures~\ref{fig:dW_D}--\ref{fig:dW_NA} plot $\Delta W_{\rm max}$ for other NA and other $D_{\rm out}$. 

\bibliography{supp}

%Manual citation list
%\begin{thebibliography}{1}
%\bibitem{Zhang:14}
%Y.~Zhang, S.~Qiao, L.~Sun, Q.~W. Shi, W.~Huang, %L.~Li, and Z.~Yang,
 % \enquote{Photoinduced active terahertz metamaterials with nanostructured
  %vanadium dioxide film deposited by sol-gel method,} Opt. Express \textbf{22},
  %11070--11078 (2014).
%\end{thebibliography}